\documentclass[11pt]{article}
\usepackage[left=2.5cm,top=2.5cm, bottom=1.2in, right=2.5cm]{geometry}

\usepackage{graphicx}
\usepackage{xcolor}
\usepackage[english]{babel}
\usepackage{amsmath}
\usepackage{amsthm}
\usepackage{amsfonts}
\usepackage{amssymb}
\usepackage{stackrel}
\usepackage{stfloats}
\usepackage{enumerate}
\usepackage[pdftex]{hyperref}

\newcommand{\bR}{\mathbb{R}}
\newcommand{\ud}{\,\mathrm{d}}
\newcommand{\ue}{\mathrm{e}}
\newcommand{\uj}{\mathrm{j}}
\renewcommand{\H}[1]{\mathrm{H}\left( {#1} \right)}
\renewcommand{\Pr}[1]{\mathrm{Pr}\left\lbrace{#1}\right\rbrace}
\newcommand{\h}[1]{\mathrm{h}\left( {#1} \right)}
\newcommand{\E}[1]{\mathbb{E}\left[{#1}\right]}
\newcommand{\Ex}[2]{\mathbb{E}_{#1} \left[ {#2} \right]}
\newcommand{\I}[2]{\mathrm{I}\left( {#1};{#2} \right)}
\newcommand{\D}[2]{\mathrm{D}\left( {#1}\|{#2} \right)}
\newcommand{\EQnum}{\addtocounter{equation}{1}\tag{\theequation}}
\newcommand{\Cov}[2]{\mathsf{Cov}\left( {#1} , {#2} \right)}

\newtheorem{theorem}{Theorem}
\newtheorem{lemma}{Lemma}

\newtheorem{corollary}{Corollary}
\newtheorem{definition}{Definition}
\newtheorem{example}{Example}

\newtheorem{remark}{Remark}

\allowdisplaybreaks
\date{}

\title{On the Capacity of a Class of Signal-Dependent Noise Channels}

\author{
	Hamid Ghourchian,
	Gholamali Aminian,\\
	Amin Gohari, 
	Mahtab Mirmohseni, and
	Masoumeh Nasiri-Kenari,\\
	\small{Department of Electrical Engineering, Sharif University of Technology, Tehran, Iran.}\\
	\small{
	E-mails: \{\mbox{h\_ghourchian}, \mbox{aminian}\}@ee.sharif.edu,
	\{\mbox{aminzadeh}, \mbox{mirmohseni}, \mbox{mnasiri}\}@sharif.edu}
	\thanks{This work was supported by INSF Research Grant on ``Nano-Network Communications". The first two authors contributed equally to this work. } 
}

\begin{document}
\maketitle

\begin{abstract}
	In some applications, the variance of additive measurement noise depends on the signal that we aim to measure.
	For instance, additive Gaussian signal-dependent noise (AGSDN) channel models are used in molecular and optical communication.
	Herein we provide lower and upper bounds on the capacity of additive signal-dependent noise (ASDN) channels.
	The idea of the first lower bound is the extension of the majorization inequality, and for the second one, it uses some calculations based on the fact that $\h{Y}>\h{Y|Z}$.
	Both of them are valid for all additive signal-dependent noise (ASDN) channels defined in the paper.
	The upper bound is based on a previous idea of the authors (``symmetric relative entropy'') and is used for the additive Gaussian signal-dependent noise (AGSDN) channels.
	These bounds indicate that   in ASDN channels  (unlike the classical AWGN channels), the capacity does not necessarily become larger by making the variance function of the noise smaller.
	We also provide sufficient conditions under which the capacity becomes infinity.
	This is complemented by a number of  conditions that imply capacity is finite and  a unique capacity achieving measure exists (in the sense of the output measure).

\if@twocolumn
\else
\fi
\end{abstract}

%%%%%%%
% Introduction
%%%%%%%
\section{Introduction} \label{sec:Introduction}
An additive Gaussian signal-dependent noise (AGSDN) channel with input $x$ and output $y$ is defined by 
\begin{equation*}
	f_{Y|X}(y|x)=\frac{1}{\sqrt{2\pi \sigma(x)^2}}
	\ue^{\frac{-(y-x)^2}{2\sigma(x)^2}},
\end{equation*}
where $\sigma(\cdot)$ is a given function from $\bR$ to $[0,\infty)$.
Alternatively, we may describe the AGSDN channel by $Y=X+\sigma(X)\cdot Z$ where $Z \sim\mathcal{N}(0,1)$ is a standard Gaussian random variable and independent of the input $X$.
For constant function $\sigma(x)=c$, the AGSDN channel reduces to a simple additive Gaussian channel.
More generally, we may relax the Gaussian assumption on $Z$ and consider an additive signal-dependent noise (ASDN) channel defined by
\begin{equation} \label{eqn:ChannelModel}
	Y=X+\sigma(X)\cdot Z,
\end{equation}
where noise $Z$ is assumed to be a continuous random variable with a given pdf $f_Z(z)$, and be independent of the input $X$.\footnote{See Definition \ref{def.ACRV} for the definition of continuous random variables.}
For instance, one can consider an ASDN with $Z$ being a truncated version of the Gaussian distribution as a better model in an application if we know that the output $Y$ has minimum and maximum values in that applications.

Below we provide a number of applications in which the ASDN channel arises. 
\begin{enumerate}
	\item The AGSDN channel appears in optical communications when modeling the shot noise or the optical amplification noise for $\sigma(x)=\sqrt{c_0^2+c_1^2x}$ \cite{moser2012capacity}.

	\item In molecular communication, the AGSDN channel with $\sigma(x)=c \sqrt{x}$ arises in the ligand receptor model, the particle sampling noise, the particle counting noise and the Poisson model for an absorbing receiver \cite{pierobon2011diffusion, aminian2016capacity, arjmandi-letter}. 
	In all cases, the reason for appearance of a Gaussian signal-dependent noise is the approximation of a binomial or Poisson distribution with a Gaussian distribution.
	Observe that the mean and variance of a binomial distribution with parameters $(n,p)$ relate to each other: the mean is $np$ and the variance is $np(1-p)$ respectively.
	As a result, the mean and variance of the approximated Gaussian distribution also relate to each other (see \cite[Section II.B]{reviewpaper} for a detailed overview). 
	
	\item Besides the above applications of ASDN in molecular communications we shall provide two other cases where this channel model is helpful:
	Consider the Brownian motion of a particle with no drift over a nonhomogeneous medium with $\sigma(x)$ denoting the diffusion coefficient of the medium at location $x$.
	The diffusion coefficient $\sigma(x)$ describes the movement variance of a particle when in location $x$.
	More specifically, the motion of the particle is described by the stochastic differential equation 
	\begin{equation*}
		\mathrm{d} X_t
		=\sigma(X_t) \ud B_t ,
	\end{equation*}
	where $B_t$ is the standard Wiener process (standard Brownian motion). Alternatively, we can express the above equation using the following It\^{o} integral 
	\begin{equation}
		X_{t+s} - X_{t}
		= \int_t^{t+s} \sigma(X_u) \ud B_u.\label{eqn:ito-formula}
	\end{equation}
	Let us denote the position of the particle at time $0$ by  $X=X_0$, and its position after $t$ seconds by $Y=X_{t}$.
	If $t$ is a small and fixed number, \eqref{eqn:ito-formula} reduces to
	\begin{equation*}
		Y=X+t\sigma(X)\cdot Z,
	\end{equation*}
	where $Z \sim\mathcal{N}(0,1)$.
	Thus, the movement of the particle follows an AGSDN channel law if $t$ is small.

	\item As another example, consider the molecular timing channel in a time-varying medium.
	In a molecular timing channel, information is encoded in the release time of molecules.
	A molecule released at time $X$ hits the receiver after a delay $Z$ at time $Y=X+Z$.
	Molecules are absorbed once they hit the receiver. As such, the distribution of $Z$ is that of the first arrival time.
	The existing literature only studies this problem when the medium is time-invariant (see  \cite{Srinivas, Khormuji11, Moser, Levy}): if the medium is uniform, time-invariant and one-dimensional, $Z$ is distributed according to the inverse Gaussian distribution (if there is a flow in the medium) or the L\'evy distribution (if there is no flow in the medium).
	As a result, the channel is called the additive inverse Gaussian noise additive channel, or the additive L\'evy noise in the literature.
	However, in a time-varying medium (or when the distance between the transmitter and receiver varies over time), the distribution of $Z$ depends on the release time $X$.
	As a result, we obtain a signal-dependent noise additive component.
	For instance, the additive noise can have a L\'evy distribution with a scale parameter that depends on input $X$.
	Using the scaling property of the L\'evy distribution, we can express this as $\sigma(X)\cdot Z$ where $Z$ is the standard L\'evy distribution, and $\sigma(X)$ is the scale parameter.
	This would be an ASDN channel.

	\item In the third item, we discussed Brownian motion  after a small time elapse.
	A Brownian motion with no drift is an example of a martingale.
	Now let us consider a martingale after a large time elapse.
	Here, the AGSDN channel also arises as a conditional distribution in any process that can be modeled by a discrete time martingale with \emph{bounded increments}.
	Assume that  $X_0, X_1, X_2, \cdots$ is such a martingale.
	Then $\E{X_n}=\E{X_0}$.
	Furthermore, by the martingale central limit theorem, the conditional distribution of $X_n$ given $X_0=x$ for large values of $n$ can be approximated by a Gaussian distribution with mean $X_0=x$ and a variance $\sigma_n(x)$ that depends on $X_0=x$.

	\item Finally, we relate the ASDN channel to real fading channels with a direct line of sight.
	Consider a scalar Gaussian fading channel
	\begin{equation} \label{eqn:fadingequation}
		Y=X+HX+N,
	\end{equation}
	where $X$ is the input, $H\sim\mathcal{N}(0, c_1)$ is the Gaussian fading coefficient and $N\sim\mathcal{N}(0, c_0)$ is the additive environment noise.
	The first $X$ term on the right-hand side of \eqref{eqn:fadingequation} corresponds to the direct line of sight, while the $HX$ term is the fading term.
	The distribution of $Y$ given $X=x$ is $\mathcal{N}(x, c_1x^2+c_0)$.
	Thus \eqref{eqn:fadingequation} can be expressed as $Y=X+\sigma(X)\cdot Z$ where
	\begin{equation*}
		\sigma(x)=\sqrt{c_1x^2+c_0},
		\qquad Z\sim \mathcal{N}(0,1).
	\end{equation*}
	A fast fading setting in which $H$ varies independently over each channel use corresponds to a memoryless ASDN channel.
\end{enumerate}

The purpose of this paper is to study the capacity of a memoryless additive signal-dependent noise (ASDN) channel defined via
\begin{equation*}
	Y=X+\sigma(X)\cdot Z,
\end{equation*}
under input cost constraints. The memoryless assumption implies that the noise $Z$ is drawn independently from $f_Z(z)$ in each channel use.

% Related Works
\textbf{Related works:}
In \cite{chan2005capacity}, vector AGSDN channels subject cost constraints are studied.
It is shown that under some assumptions, the capacity achieving distribution is a discrete distribution.
The  AGSDN channel with $\sigma(x)=\sqrt{c_0^2+c_1^2x}$ is investigated in \cite{moser2012capacity} wherein capacity upper and lower bounds are derived considering peak and average constraints. 

Note that the memoryless AGSDN includes the additive white Gaussian noise (AWGN) channel as its special case.
The capacity of AWGN channel under power constraint is classical and is obtained by an input of Gaussian random variable.
Its capacity under both average and peak power constraints is quite different, as the capacity achieving input distribution is discrete with a finite number of mass points  \cite{smith1971information}.
See \cite{jiang2016tight, lapidoth2009capacity} for further results on the capacity of the AWGN channel with both average and peak power constraints. 

% Our Contribution
\textbf{Our contributions:} 
Our contributions in this work can be summarized as follows:
\begin{itemize}
	\item We provide a new tool for bounding the capacity of continuous input/output channels. Note that
	\begin{equation*}
		\I{X}{Y}=\h{Y}-\h{Y|X}.
	\end{equation*}
	We provide two sufficient conditions under which $\h{Y}\geq \h{X}$, which results in
	\begin{equation*}
		\I{X}{Y}\geq \h{X}-\h{Y|X},
	\end{equation*}
	and leads to lower bounds on the channel capacity of an ASDN channel.

	\item It is known that increasing the noise variance of an AWGN channel decreases its capacity.
	However, we show that this is no longer the case for signal-dependent noise channels:
	the constraint $\sigma_1(x)\geq \sigma_2(x)$ for all $x$ does not necessarily imply that the capacity of an AGSDN channel with $\sigma_1(x)$ is less than or equal to the capacity of an AGSDN with $\sigma_2(x)$.

 	\item We identify conditions under which the capacity of the ASDN channel becomes infinity.
 	In particular, this implies that the capacity of a AGSDN channel with
	\begin{equation*}
		\sigma(x)=\sqrt{c_1x^2+c_0}
	\end{equation*}
	tends to infinity as $c_0$ tends to zero.
	Thus, the capacity of the real Gaussian fast fading channel given earlier in this section tends to infinity as $c_0$ tends to zero.
	This parallels a similar result given in \cite{Chen04} for complex Gaussian fading channels.

	\item We provide a new upper bound for the AGSDN channel based on the KL symmetrized upper bound of \cite{aminian2015capacity}.
	This upper bound is suitable for the low SNR regime, when $\sigma(x)$ is large. This is in contrast with the upper bound of \cite[Theorems 4, 5]{moser2012capacity} for AGSDN channels with  $\sigma(x)=\sqrt{c_0^2+c_1^2x}$ which is suitable for large values of peak and average constraints.  Furthermore, we give our upper bound for a large class of functions $\sigma(x)$ while the technique of \cite{moser2012capacity} is tuned for $\sigma(x)=\sqrt{c_0^2+c_1^2x}$. 
\end{itemize}

% Paper Organization
This paper is organized as follows.
Section \ref{sec:DefNot} includes some of primary definitions and notations.
In Section \ref{UpperAndLowerBounds}, our main results are given.
This includes two lower bounds and one upper bound on the capacity of the ASDN channel.
There are some useful lemmas in Section \ref{sec:UsefulLemma} used in the paper.
The numerical results and plots are given in Section \ref{Numericalresults}.
The proofs of our results are given in Section \ref{sec:Proofs}.
%and the bounds are plotted in  Section \ref{Numericalresults}.

%%%%%%%%%%%
% Definition & Notations
%%%%%%%%%%%
\section{Definitions and Notations} \label{sec:DefNot}
In this section we review the definitions of continuous and discrete random variables, as well as entropy and differential entropy, relative entropy and mutual information.

% Notations
Throughout this paper all the logarithms are in base $\ue$.
Random variables are denoted by capital letters, and probability measure functions are denoted by letter $\mu$.
The collection of Borel measurable sets in $\bR$ is denoted by $\mathcal{B}(\bR)$.
We sometimes use a.e. and $\mu$-a.e. as a short-hand for ``almost everywhere" and ``$\mu$-almost everywhere", respectively.
The set $\mathcal{A}$ is $\mu$-a.e., when
\begin{equation*}
	\int_\mathcal{A}{\ud \mu} = 0.
\end{equation*}
The set $\mathcal{A}$ is a.e. if it is $\mu$-a.e. when $\mu$ is the Lebesgue measure.

% Definition: Relative Entropy
\begin{definition}[Relative Entropy] \cite[Section 1.4]{Ihara93} \label{def:DKL}
	For random variables $X$ and $Y$ with probability measures $\mu_X$ and $\mu_Y$, the relative entropy between $X$ and $Y$ is defined as follows:
	\begin{equation*}
		\D{\mu_X}{\mu_Y}
		=\D{X}{Y} :=
		\begin{cases}
			\E{\log\frac{\ud\mu_X}{\ud\mu_Y}(X)} & \mu_X \ll \mu_Y \\
			+\infty & \text{o.w.}
		\end{cases},
	\end{equation*}
	where $\frac{\ud\mu_X}{\ud\mu_Y}$ is the Radon-Nikodym derivative and $\mu_X \ll \mu_Y$ means $\mu_X$ is absolutely continuous \emph{w.r.t.} $\mu_Y$ \emph{i.e.} $\mu_X(\mathcal{A})=0$ for all $\mathcal{A}\in\mathcal{B}$ if $\mu_Y(\mathcal{A})=0$, where $\mathcal{B}$ is the Borel $\sigma$-field of the space over which the measures are defined.
\end{definition}

% Definition: Mutual Information
\begin{definition}[Mutual Information] \cite[Section 1.6]{Ihara93}
	For random variables $X,Y$ with joint probability measure $\mu_{X,Y}$, the mutual information between $X$ and $Y$ is defined as follows:
	\begin{equation*}
		\I{X}{Y}=\D{\mu_{X,Y}}{\mu_{X}\mu_{Y}},
	\end{equation*}
	where $\mu_{X}\mu_{Y}$ is the product measure defined as
	\begin{equation*}
		(\mu_{X}\mu_{Y})(\mathcal{A}, \mathcal{C})
		=\mu_X(\mathcal{A})\mu_Y(\mathcal{C}),
	\end{equation*}
	where $\mathcal{A}\in\mathcal{B}_X$ the Borel $\sigma$-field of the space over which $\mu_X$ is defined, and $\mathcal{C}\in\mathcal{B}_Y$ the Borel $\sigma$-field of the space over which $\mu_Y$ is defined.
	
	Similarly, for three random variable $X,Y,Z$ with joint measure $\mu_{X,Y,Z}$, conditional mutual information $\I{X}{Y|Z}$ is defined as $\I{X}{Y,Z}-\I{X}{Z}$.
\end{definition}

% Definition: Continuous Random Variable
\begin{definition} [Continuous Random Variable] \label{def.ACRV} \cite{chan2005capacity}
	Let $X$ be a real-valued and random variable that is measurable with respect to $\mathcal{B}(\bR)$.
	We call $X$ a \emph{continuous} random variable if its probability measure $\mu_X$, induced on $(\bR,\mathcal{B})$, is absolutely continuous with respect to the Lebesgue measure for $\mathcal{B}(\bR)$ (i.e., $\mu(\mathcal{A})=0$ for all $\mathcal{A}\in\mathcal{B}$ with zero Lebesgue measure).
	We denote the set of all absolutely continuous probability measures by $\mathcal{AC}$.
	Note that the Radon-Nikodym theorem implies that for each random variable $X$ with measure $\mu_X\in\mathcal{AC}$ there exists a $\mathcal{B}(\bR)$-measurable function $f_X:\mathbb{R}\to [0,\infty)$, such that for all $\mathcal{A}\in\mathcal{B}(\bR)$ we have that
	\begin{equation}
		\mu_X(\mathcal{A})
		=\Pr{X\in \mathcal{A}}
		=\int_\mathcal{A}{f_X(x) \ud x}.
	\end{equation}
	The function $f_X$ is called the probability density function (pdf) of $X$ \cite[p. 21]{Ihara93}.
	We denote pdf of absolutely continuous probability measures by letter $f$.
	%The property $\mu_X\in\mathcal{AC}$ is alternatively written as $X\in\mathcal{AC}$.
\end{definition}

% Definition: Discrete Random Variable
\begin{definition}[Discrete Random Variable] \cite{chan2005capacity} \label{def.ADRV}
	A random variable $X$ is discrete if it takes values in a countable alphabet set $\mathcal{X}\subset\mathbb{R}$.
\end{definition}

Probability mass function (pmf) for discrete random variable $X$ with probability measure $\mu_X$ is denoted by $p_X$ and defined as follows:
\begin{equation*}
	p_X(x) := \mu_X(\{x\}) = \Pr{X=x},
	\qquad\forall x\in\mathcal{X}.
\end{equation*}

% Definition: Entropy and Differential Entropy
\begin{definition}[Entropy and Differential Entropy] \label{def:Ent} \cite[Chapter 2]{Cover06}
	We define entropy $\H{X}$, for a discrete random variable $X$ with measure $\mu_X$ and pmf $p_X$ as
	\begin{equation*}
		\H{X}
		=\H{\mu_X}=\H{p_X}
		:= \sum_{x}p_X(x)\log\frac{1}{p_X(x)},
	\end{equation*}
	if the summation converges.
	Observe that
	\begin{equation*}
		\H{X}=\E{\log\frac{1}{p_X(X)}}.
	\end{equation*}
	For a continuous random variable $X$ with measure $\mu_X$ and pdf $f_X$, we define differential entropy $\h{X}$ as
	\begin{equation*}
		\h{X}=\h{\mu_X}=\h{p_X}
		:= \int_{-\infty}^{+\infty}f_X(x)\log\frac{1}{f_X(x)},
	\end{equation*}
	if the integral converges.
	Similarly, the differential entropy is the same as
	\begin{equation*}
		\h{X}=\E{\log\frac{1}{f_X(X)}}.
	\end{equation*}
	
	Similarly, for two random variables $X,Y$, with measure $\mu_{X,Y}$, if for all $x$, $\mu_{Y|X}(\cdot|x)$ is absolutely discrete with pmf $p_{Y|X}(\cdot|x)$, the conditional entropy $\H{Y|X}$ is defined as
	\begin{equation*}
		\H{Y|X} = \E{\log\frac{1}{p_{Y|X}(Y|X)}}.
	\end{equation*}
	Likewise, for two random variables $X,Y$, with measure $\mu_{X,Y}$, if for all $x$, $\mu_{Y|X}(\cdot|x)$ is absolutely continuous with pdf $f_{Y|X}(\cdot|x)$, the conditional differential entropy $\h{Y|X}$ is defined as
	\begin{equation*}
		\h{Y|X} = \E{\log\frac{1}{f_{Y|X}(Y|X)}}.
	\end{equation*}
\end{definition}

	We allow for differential entropy to be $+\infty$ or $-\infty$ if the integral is convergent to $+\infty$ or $-\infty$, \emph{i.e.}, we say that 
	\begin{equation*}
		\h{X} = +\infty,
	\end{equation*}
	if and only if
	\begin{align*}
		&\int_\mathcal{A^+}{f_X(x) \log\frac{1}{f_X(x)} \ud x} = +\infty, \text{ and}\\
		&\int_\mathcal{A^-}{f_X(x) \log\frac{1}{f_X(x)} \ud x} \text{ converges to a finite number}
	\end{align*}
	where
	\begin{equation*}
		\mathcal{A}^+=\{x:f_X(x) \leq 1\},
		\qquad\mathcal{A}^-=\{x:f_X(x) > 1\}.
	\end{equation*}
	Similarly, we define $\h{X} = -\infty$. When we write that $\h{X}>-\infty$, we mean that the differential entropy of $X$ exists and is not equal to $-\infty$.
The following example, from \cite{Chen04}, demonstrates the differential entropy can be $+\infty$ or $-\infty$.

% Example: h=+inf
\begin{example}  \label{exp:h=+-inf} Differential entropy becomes plus infinity for the following pdf defined over $\bR$ \cite{Chen04}:
	\begin{equation*}
		f(x)=
		\begin{cases}
			\frac{1}{x(\log x)^2}, & x>\ue, \\
			0, & x \leq \ue.
		\end{cases}\label{def-p-eq}
	\end{equation*}
On the other hand, as shown in \cite{Chen04}, differential entropy is minus infinity for 
	\begin{equation*}
		g(x)=
		\begin{cases}
			\frac{-1}{x\log x (\log(-\log x))^2}, & 0<x<\ue^{-\ue}, \\
			0, & \text{otherwise}.
		\end{cases}
	\end{equation*}
\end{example}

% Definition: Regular Functions
\begin{definition}[Riemann integrable functions] \label{def:RegFun} Given $-\infty\leq\ell < u\leq +\infty$, in this work, we utilize Riemann integrable functions $g:(\ell,u)\mapsto\bR$ on open interval $(\ell,u)$. Such functions satisfy the property that for any $c\in(\ell,u)$, the function 
	\begin{equation*}
		h(x) = \int_c^x{g(t) \ud t},
	\end{equation*}
is well-defined. By the fundamental theorem of calculus, $h(\cdot)$ is continuous on $(\ell,u)$ (but not necessarily differentiable unless $g$ is continuous).
\end{definition}
As an example, consider the function $g(x)=1/x$ for $x\neq 0$, and $g(0)=0$ otherwise. This function is Riemann integrable on the restricted domain $(0, \infty)$, but not integrable on $(-1,1)$.

%%%%%%%
% Main Results
%%%%%%%
\section{Main Results}\label{UpperAndLowerBounds}
% Channel Capacity Definition
We are interested in the capacity of an ASDN channel with the input $X$ taking values in a set $\mathcal{X}$ and satisfying the cost constraint $\mathbb{E}[g_i(X)]\leq 0,~\forall i=1,2,\cdots, k$ for some functions $g_i(\cdot)$. The common power constraint corresponds to  $g_i(x)=x^2-p$ for some $p\geq 0$, but we allow for more general constraints. Then, given a density function $f_Z(z)$ for the noise $Z$ and function $\sigma(\cdot)$, we consider the following optimization problem: 
\begin{equation} \label{eqn:capacity-def}
	C=\sup_{\mu_X\in \mathcal{F}}{\I{X}{Y}},
\end{equation}
where $X$ and $Y$ are related via \eqref{eqn:ChannelModel} and 
\if@twocolumn
\begin{align*}
	\mathcal{F}
	=\{\mu_X\big|
	&\textnormal{supp}(\mu_X) \subseteq \mathcal{X}, \\
	&\mathbb{E}[g_i(X)]\leq 0 \text{ for all } i=1,\cdots, k\}.\EQnum\label{eqn:FeasibleSet fx}
\end{align*}
\else
\begin{equation} \label{eqn:FeasibleSet fx}
	\mathcal{F}
	=\{\mu_X\big|
	\textnormal{supp}(\mu_X) \subseteq \mathcal{X},
	\mathbb{E}[g_i(X)]\leq 0 \text{ for all } i=1,\cdots, k\}.
\end{equation}
\fi
We sometimes use $\textnormal{supp}(X)$ to denote the support of measure $\mu_X$, $\textnormal{supp}(\mu_X)$, when the probability measure on $X$ is clear from the context.

As an example, if, in an application, input $X$ satisfies $\ell\leq X\leq u$, the set $\mathcal{X}$ can be taken to be $[\ell, u]$ to reflect this fact; similarly, the constraint $0<X\leq u$ reduces to $\mathcal{X}=(0, u]$, and $0\leq \ell\leq |X|\leq u$ reduces to $\mathcal{X}=[-u,-\ell]\cup [\ell, u]$.

The rest of this section is organized as follows: in Section \ref{sec:Capacity}, we provide conditions that imply finiteness of the capacity of an ASDN channel.
In Section \ref{subsec:LowBnd}, we review the ideas used for obtaining lower bounds in previous works and also in this work. Then, based on the new ideas introduced in this work, we provide two different lower bounds  in Sections \ref{sec:MajLB} and \ref{sec:1/f+logCapLB}.
Finally, in Section \ref{subsec:KLUpBnd}, we provide an upper bound for AGSDN channels.

%%%%%%%%%%%%%%%%
% Existence & Finiteness of Capacity
\subsection{Existence and Finiteness of Channel Capacity} \label{sec:Capacity}

% Theorem: Capacity Achievability
\begin{theorem} \label{thm:CapDef}
	Assume that an ASDN channel satisfies the following properties:
\begin{itemize}
\item $\mathcal{X}$ is a closed and also bounded subset of $\mathbb{R}$, \emph{i.e.}, there exists $u\geq 0$ such that $\mathcal{X}\subseteq [-u,u]$;
\item Real numbers $0<\sigma_\ell<\sigma_u$ exist such that $\sigma_\ell\leq \sigma(x)\leq \sigma_u$ for all $x\in\mathcal{X}$;
\item Positive real $m$ and $\gamma$ exist such that $f_Z(z)\leq m<\infty$ \emph{(a.e.)}, and $\E{|Z|^\gamma}=\alpha < \infty$;
\item The cost constraint functions $g_i(\cdot)$ are bounded over $\mathcal{X}$. 
\end{itemize}
Then, the capacity of the ASDN channel is finite.
	Furthermore there is a capacity achieving probability measure; in other words, the capacity $C$ can be expressed as a maximum rather than a supremum:
	\begin{equation*}
		C=\max_{\mu_X\in \mathcal{F}}{\I{X}{Y}}.
	\end{equation*}
Moreover, the output distribution is unique, \emph{i.e.} if $\mu_{X_1}$ and $\mu_{X_2}$ both achieves the capacity, then
	\begin{equation*}
		f_{Y_1}(y) = f_{Y_2}(y),
		\qquad\forall y\in\bR,
	\end{equation*}
	where $f_{Y_1}$ and $f_{Y_2}$ are the pdfs of the output of the channel when the input probability measures are $\mu_{X_1}$ and $\mu_{X_2}$, respectively.
\end{theorem}

% Remark: The idea of the proof is old
\begin{remark}
	The above theorem is a generalization of that given in \cite[Theorem 1]{chan2005capacity} for the special case of Gaussian noise $Z$.
\end{remark}

The proof can be found in Section \ref{subsec:prf:thm:CapDef}.
To give a partial converse of the above theorem, consider the case that the second assumption of the above theorem fails, \emph{i.e.,} when there is a sequence $\{x_i\}$ of elements in $\mathcal X$ such that $\sigma(x_i)$ converges to zero or infinity.
The following theorem shows that input/output mutual information can be infinity in such cases.

% Theorem: Necessary & Sufficient Condition for C=inf
\begin{theorem} \label{thm:iffC=inf}
	Consider an ASDN channel with $\sigma:\mathcal{X}\mapsto[0,+\infty)$ where $\mathcal X$ is not necessarily a closed set.
	Suppose one can find  a sequence $\{\tilde{x}_i\}$ of elements in $\mathcal X$ such that $\sigma(\tilde x_i)$ converges to $0$ or $+\infty$ such that 
	\begin{itemize}
	\item As a sequence on real numbers, $\{\tilde{x}_i\}$ has a limit (possibly outside $\mathcal{X}$), which we denote by $c$. The limit $c$ can be plus or minus infinity.
	\item One can find another real number $c'\neq c$ such that the open interval $\mathcal{E}=(c, c')$ (or $\mathcal{E}=(c',c)$ depending on whether $c'>c$ or $c'<c$) belongs to $\mathcal{X}$.
	Furthermore, $\tilde{x}_i\in \mathcal{E}$, and $\sigma(\cdot)$ is monotone and continuous over $\mathcal{E}$.
	\footnote{We only require  monotonicity here, and not strictly monotonicity.}
\end{itemize}
	Then one can find a measure $\mu_X$ defined on $\mathcal{E}$ such that $\I{X}{Y}=\infty$ provided that $Z$ is a continuous random variable and has the following regularity conditions:
\if@twocolumn
	\begin{align*}
		&|\h{Z}|<\infty,\\
		&\exists\delta>0:\Pr{Z>\delta},\Pr{Z<-\delta}>0,
	\end{align*}
\else
	\begin{equation*}
		|\h{Z}|<\infty,
		\qquad\exists\delta>0:\Pr{Z>\delta},\Pr{Z<-\delta}>0,
	\end{equation*}
\fi
	Furthermore, there is more than one measure $\mu_X$ that makes $\I{X}{Y}=\infty$.
	In fact, input $X$ can be both a continuous or discrete random variable, \emph{i.e.,} one can find both an absolutely continuous measure with pdf $f_{X}$ and discrete pmf $p_X$ such that  $\I{X}{Y}$ is infinity when the measure on input is either  $f_{X}$ or $p_X$.
\end{theorem}
The proof can be found in Section \ref{subsec:prf:thm:iffC=inf} and uses some of the results that we prove later in the paper.

% Remark: f_{\sigma}(x)=x^a
\begin{remark}
	As an example, consider an AGSDN channel with $\mathcal{X}=(0,u)$ for an arbitrary $u>0$, and $\sigma(x)=x^\alpha$ for $\alpha\neq0$.
	For this channel, we have $C=+\infty$ if we have no input cost constraints. Setting $\alpha=1$, this shows that the capacity of the fast-fading channel given in \eqref{eqn:fadingequation} is infinity if $c_0=0$; that is when there is no  additive noise.
	This parallels a similar result given in \cite{Chen04} for complex Gaussian fading channels.
\end{remark}

% Remark: sigma1>sigma2 -/-> C1 < C2
\begin{remark}
	It is known that increasing the noise variance of an AWGN channel decreases its capacity.
However, we show that this is no longer the case for signal-dependent noise channels:
	Consider two AGSDN channels with parameters $\sigma_1(x)$ and $\sigma_2(x)$, respectively, which are defined over $\mathcal{X}=(0,1)$ with the following formulas:
	\begin{equation*}
		\sigma_1(x) = 1,
		\qquad \sigma_2(x) = \frac{1}{x}.
	\end{equation*}
No input cost constraints are imposed. 
	It is clear that $\sigma_2(x) > \sigma_1(x)$ for all $x\in\mathcal{X}$.
	However, by considering the constraint $0<X<1$, from Theorem \ref{thm:CapDef} we obtain that the capacity of the first channel is finite, while from Theorem \ref{thm:iffC=inf}, we obtain that the capacity of the second channel is $\infty$. Therefore, the constraint $\sigma_1(x) > \sigma_2(x)$ for all $x\in\mathcal{X}$ does not necessarily imply that the capacity of an AGSDN channel with $\sigma_1(x)$ is less than or equal to the capacity of an AGSDN with $\sigma_2(x)$.
\end{remark}

%%%%%%%%
% Lower Bounds
\subsection{Lower Bounds on Capacity} \label{subsec:LowBnd}
To compute capacity from \eqref{eqn:capacity-def}, one has to take maximum over probability measures in a potentially large class $\mathcal{F}$.
Practically speaking, one can only find a finite number of measures $\mu_1, \mu_2, \cdots, \mu_k$ in $\mathcal{F}$ and evaluate input/output mutual information for them.
Ideally, $\{\mu_i\}$ should form an $\epsilon$-covering of the entire $\mathcal{F}$ (with an appropriate distance metric), so that mutual information at every arbitrary measure in $\mathcal{F}$ can be approximated with one of the measures $\mu_i$.
This can be computationally cumbersome, even for measures defined on a finite interval.
As a result, it is desirable to find explicit lower bounds on the capacity.
Observe that $\I{X}{Y}=\h{Y}-\h{Y|X}$.
To compute the term $\h{Y|X}$, observe that given $X=x$, we have $Y=x+\sigma(x)\cdot Z$ and thus $\h{Y|X=x}= \log \sigma(x)+\h{Z}$ (see Lemma \ref{lmm:hY|X}).
Thus,
\begin{equation*}
	\h{Y|X} = \E{\log \sigma(X)} + \h{Z}.
\end{equation*}
However, the term $\h{Y}$ is more challenging to handle.
Authors in \cite{moser2012capacity} consider an AGSDN channel with $\sigma(x)=\sqrt{c_0^2+c_1^2x}$ for $x\geq 0$, as well as show that $\h{Y}\geq \h{X}$ and hence $\I{X}{Y}\geq \h{X}-\h{Y|X}$.
This implies that  instead of maximizing $\I{X}{Y}$, one can maximize $\h{X}-\h{Y|X}$ to obtain a lower bound. 

The proof of the relation $\h{Y}\geq \h{X}$ in \cite{moser2012capacity} is non-trivial; we review it here to motivate our own techniques in this paper.
First consider the special case of $c_1=0$.
In this case, we get $\sigma(x)=c_0$ and the AGDSN reduces to AWGN channel $Y=X+Z$.
In this special case, one obtains the desired equation by writing
\begin{equation} \label{eqn:hYgreaterhX1}
	\h{Y}\geq \h{Y|Z}=\h{X+Z|Z}=\h{X|Z}=\h{X}.
\end{equation}
However, the above argument does not extend for the case of $c_1>0$ since  $\sigma(x)=\sqrt{c_0^2+c_1^2x}$ depends on $x$.
As argued in \cite{moser2012capacity}, without loss of generality, one may assume that $c_0=0$; this is because one can express a signal-dependent noise channel with $\sigma(x)=\sqrt{c_0^2+c_1^2x}$ as
\begin{equation*}
	Y=X+c_1 \sqrt{X} Z_1+c_0 Z_0,
\end{equation*}
where $Z_0$ and $Z_1$ are independent standard normal variables.
Thus, we can write $Y=Y_1+c_0 Z_0$ where $Y_1=X+c_1 \sqrt{X} Z_1$.
From the argument for AWGN channels, we have that $\h{Y}\geq \h{Y_1}$.
Thus, it suffices to show that $\h{Y_1}\geq\h{X}$. This is the special case of the problem for $c_0=0$ and corresponds to $\sigma(x)=c_1\sqrt{x}$. 

To show $\h{Y}\geq\h{X}$ when $Y=X+c_1 \sqrt{X} Z$, more advanced ideas are utilized in \cite{moser2012capacity}.
The key observation is the following: assume that
\begin{equation*}
	X\sim g_X(x)=\frac{1}{\alpha}e^{-\frac{x}{\alpha}}\mathbf{1}[x\geq 0],
\end{equation*}
be exponentially distributed with mean $\E{X}=\alpha$.
Then $Y$ has density
\begin{equation*}
	g_Y(y)
	= \frac{1}{\sqrt{\alpha(\alpha+2 c_1^2)}}
	\exp\left( \frac{\sqrt{\alpha}y-\sqrt{\alpha+2 c_1^2}|y|}{\sqrt{\alpha}c_1^2}\right).
\end{equation*}
Then, for any arbitrary input distribution $f_X$, from the data processing property of the relative entropy, we have
\begin{equation*}
	\D{f_Y}{g_Y} \leq \D{f_X}{g_X}
\end{equation*}
where $f_Y$ is the output density for input density $f_X$.
Once simplified, this equation leads to $\h{f_Y}\geq \h{f_X}$.

The above argument crucially depends on the particular form of the output distribution corresponding to the input exponential distribution.
It is a specific argument that works for the specific choice of $\sigma(x)=\sqrt{c_0^2+c_1^2x}$ and normal distribution for $Z$, and cannot be readily extended to other choices of $\sigma(\cdot)$ and $f_Z(z)$.
In this paper, we propose two approaches to handle more general settings:

\begin{itemize}
	\item (Idea 1:) We provide the following novel general lemma that establishes $\h{Y}\geq \h{X}$ for a large class of ASDN channels. 
% Lemma: Majorization h(Y)>h(X)
	\begin{lemma} \label{thm:Maj hY>hX} 
		Take an arbitrary channel characterized by the conditional pdf $f_{Y|X}(\cdot|x)$ satisfying 
		\begin{equation}
			\int_\mathcal{X}{f_{Y|X}(y|x) \ud x} \leq 1,
			\qquad\forall y\in\mathcal{Y},\label{eqn:maj1}
		\end{equation}
		where $\mathcal{X}$ and $\mathcal{Y}$ are the support of channel input $X$ and channel output $Y$, respectively.
		Take an arbitrary input pdf $f_X(x)$ on $\mathcal{X}$ resulting in an output pdf $f_Y(y)$ on $\mathcal{Y}$.
		Assuming that $\h{X}$ and $\h{Y}$ exist, we have
		\begin{equation*}
			\h{Y} \geq \h{X}
		\end{equation*}
	\end{lemma}
	The proof is provided in Section \ref{subsec:Proof:thm:Maj hY>hX}. 

	As an example, Lemma \ref{thm:Maj hY>hX} yields an alternative proof for the result of \cite{moser2012capacity} for an AGSDN channel.
	Note that, as we mentioned before, in order to prove that $\h{Y} \geq \h{X}$ for $\sigma(x)=\sqrt{{c'}^2+c^2 x}$, we only need to prove it for $\sigma(x)=c\sqrt{x}$.
	To this end, observe that since $\mathcal{X}\subseteq [0,+\infty)$, we have
\if@twocolumn
	\begin{align*}
		&\int_{0}^{\infty}
		{\frac{1}{\sqrt{2\pi c^2 x)}}
		\ue^{\frac{-(y-x)^2}{2 c^2 x}} \ud x}\\
		&=\int_{\hat{c}}^{\infty}
		{\frac{1}{\sqrt{2\pi c_1^2 w}} \ue^{\frac{-(y-w)^2}{2 c_1^2 w}} \ud w} \\
		&\leq \int_{0}^{\infty}
		{\frac{1}{\sqrt{2\pi c_1^2 w}} \ue^{\frac{-(y-w)^2}{2 c_1^2 w}} \ud w}\\
		&=\int_{0}^{\infty}
		{\frac{\sqrt{2}}{\sqrt{\pi c_1^2 }} \ue^{\frac{-(y-v^2)^2}{2 c_1^2 v^2}} \ud v}\nonumber\\
		&=\ue^{\frac{y-|y|}{c_1^2}} \EQnum\label{eqn:integral-erf} \\
		&=
		\begin{cases}
			1 \quad \textit{if}\quad y\geq 0, \\
			\ue^{\frac{2y}{c_1^2}}<1 \quad \textit{if}\quad y<0,
		\end{cases}
	\end{align*}
\else
	\begin{align*}
		\int_\mathcal{X}{f_{Y|X}(y|x) \ud x}
		\leq & \int_{0}^{\infty}
		{\frac{1}{\sqrt{2\pi c^2 x}}
		\ue^{-\frac{(y-x)^2}{2 c^2 x}} \ud x} \\
		=&\int_{0}^{\infty}
		{\frac{\sqrt{2}}{\sqrt{\pi c^2 }} \ue^{\frac{-(y-v^2)^2}{2 c^2 v^2}} \ud v} \\
		=&
		\begin{cases}
			1 & y\geq 0 \\
			\ue^{\frac{2y}{c_1^2}} & y<0
		\end{cases} 
		\quad\leq 1. \EQnum\label{eqn:integral-erf}
	\end{align*}
\fi
	where $x=v^2$, and $v\geq 0$.
	The proof for equation \eqref{eqn:integral-erf} is given in Appendix \ref{Appendix:integral-erf}. 

	\item (Idea 2:) We provide a variation of the type of argument given in \eqref{eqn:hYgreaterhX1} by introducing a number of new steps. This would adapt the argument to ASDN channels.
\end{itemize}

In the following sections, we discuss the above two ideas separately. 

%%%%%%%%%%%%%%
% Lower Bound: Majorization
\subsection{First Idea for Lower Bound} \label{sec:MajLB}

% Theorem: Majorization Capacity Lower Bound
\begin{theorem} \label{thm:MajCapLB}
	Assume an ASDN channel defined in \eqref{eqn:ChannelModel}, where $\sigma:(\ell,u)\mapsto (0,+\infty)$ with $-\infty\leq\ell<u\leq+\infty$, and noise with pdf $f_Z(z)$ such that
	\begin{equation} \label{eqn:MajCapLB:p(y|x)const}
		\int_\ell^u{\frac{1}{\sigma(x)} f_Z\left(\frac{y-x}{\sigma(x)}\right) \ud x}
		\leq 1, \qquad \forall y,
	\end{equation}
	\begin{equation} \label{eqn:MajCapLB:1/fconst}
		\frac{1}{\sigma(x)} \text{ is Riemann integrable on } (\ell,u).
	\end{equation}
	Then, if $X$ is continuous random variable with pdf $f_X(x)$ supported over $(\ell , u)$,
	\begin{equation*}
		\I{X}{Y} \geq \h{\varphi(X)}-\h{Z},
	\end{equation*}
	provided that the integrals defining  $\h{\varphi(X)}$ and $\h{Z}$ converge to a real number or $\pm\infty$.
	The function $\varphi(x)$ is an increasing function of $x$ defined by
	\begin{align} \label{eqn:phi=1/f}
		\varphi(x) = \int_c^x{\frac{1}{\sigma(t)}\ud t},
		\qquad\forall x\in (\ell,u),
	\end{align}
	where $c\in(\ell,u)$ is arbitrary.
\end{theorem}

% Remark
\begin{remark}\label{rmk3}
	Note that for any  $c\in(\ell,u)$, $\varphi(x)$ is well defined (see Definition \ref{def:RegFun}). 
	By selecting a different $c'\in(\ell,u)$ we obtain a different function $\varphi'(x)$ such that
	\begin{equation*}
		\varphi'(x)-\varphi(x) = \int_{c'}^{c}{\frac{1}{\sigma(t)}\ud t}<\infty.
	\end{equation*}
	However, $\h{\varphi(X)}$ is invariant with respect to adding constant terms, and thus invariant with respect to different choices of $c\in(\ell,u)$.
\end{remark}

The above theorem is proved in Section \ref{subsec:Proof:thm:MajCapLB}.

% Corollary: More Constraints
\begin{corollary} \label{cor:MajCapLB:MorConst}
	Let $W=\varphi(X)$. Since $\varphi(\cdot)$ is a one-to-one function (as $\sigma(x)>0$), we obtain 
	\begin{equation*}
		\max_{\mu_X\in\mathcal{F}\cap \mathcal{AC}}\h{\varphi(X)} -\h{Z}=\max_{f_W\in\mathcal{G}}\h{W} -\h{Z},
	\end{equation*}
where $\mathcal{F}$ is defined in \eqref{eqn:FeasibleSet fx}, and $W \sim f_W$ belongs to
\if@twocolumn
	\begin{align*}
		\mathcal{G}
		=\{f_W(\cdot)\big| &\mu_W \in \mathcal{AC},\\
		&\textnormal{supp}(\mu_W) \subseteq \varphi(\mathcal{X}), \\
		&\mathbb{E}[g_i(\varphi^{-1}(W))]\leq 0 \text{ for all } i=1,\cdots, k\}.
	\end{align*}
\else
	\begin{equation*}
		\mathcal{G}
		=\{f_W(\cdot)\big| \mu_W \in \mathcal{AC},
		~\textnormal{supp}(\mu_W) \subseteq \varphi(\mathcal{X}),
		~\mathbb{E}[g_i(\varphi^{-1}(W))]\leq 0 \text{ for all } i=1,\cdots, k\}.
	\end{equation*}
\fi
Here $\varphi(\mathcal{X})=\{\varphi(x): x\in\mathcal{X}\}$.
Hence, from Theorem \ref{thm:MajCapLB} we obtain that
	\begin{equation*}
		\max_{\mu_X\in\mathcal{F}}\I{X}{Y}
		\geq\max_{f_W\in\mathcal{G}}\h{W} -\h{Z}.
	\end{equation*}
\end{corollary}
In order to find the maximum of $\h{W}$ over $f_W\in\mathcal{G}$, we can use known results on maximum entropy probability distributions, \emph{e.g.,} see \cite[Chapter 3.1]{Ihara93}.

% Corollary: Finite Support
\begin{corollary}\label{cor:MajCapLB:FiniteSupport}
	Consider an ASDN channel satisfying \eqref{eqn:MajCapLB:p(y|x)const} and \eqref{eqn:MajCapLB:1/fconst}.
	Assume that the only input constraint is $\mathcal{X}=(\ell,u)$ \emph{i.e.} $\ell < X < u$.
	Then, from Corollary \ref{cor:MajCapLB:MorConst}, we obtain the lower bound 
	\begin{equation*}
		\max_{f_W\in\mathcal{G}}\h{W} -\h{Z}
		=\log{\left(\int_\ell^u{\frac{1}{\sigma(x)}\ud x}\right)} - \h{Z},
	\end{equation*}
	by taking a uniform distribution for $f_W(w)$ over $\varphi(\mathcal{X})$ if this set is bounded \cite[Section 3.1]{Ihara93}.
	Else, if $\varphi(\mathcal{X})$ has an infinite length, the capacity is infinity by choosing a pdf for $W$ whose differential entropy is infinity (see Example \ref{exp:h=+-inf}).
	The equivalent pdf $f_X(x)$ for $X$ is the pdf of $\varphi^{-1}(W)$.
\end{corollary}

For more insight, we provide the following example.
% Example: AWGN Major LB
\begin{example} \label{exp:AWGNMajLB}
	Consider an AWGN channel (namely, an AGSDN channel with $\sigma(x)=\sigma_0$) with $\mathcal{X}=\mathbb{R}$ and $Z\sim\mathcal{N}(0,1)$. Let us restrict to measures that satisfy the power constraint $\E{X^2}\leq P$; that is $g_1(x)=x^2$.
	Since
	\begin{align*}
		\int_\bR{\frac{1}{\sigma(x)} f_Z\left(\frac{y-x}{\sigma(x)}\right) \ud x}
		= \int_\bR{\frac{1}{\sqrt{2\pi\sigma_0^2}}\ue^{-\frac{(y-x)^2}{2\sigma_0^2}} \ud x}
		=1,
	\end{align*}
we can apply Corollary \ref{cor:MajCapLB:MorConst}. Here $W=\varphi(X)=x/\sigma_0$; thus, the lower bound is
	\begin{equation}
		C\geq \max_{f_W(\cdot):\E{W^2}\leq\frac{P}{\sigma_0^2}}{\h{W} -\h{Z}}
		=\frac{1}{2}\log{\frac{P}{\sigma_0^2}},\label{eqnAWGNlower}
	\end{equation}
	where it is achieved by Gaussian distribution $W\sim\mathcal{N}(0,\sqrt{P}/\sigma_0)$\cite[Section 12.1]{Cover06}.
	It is well-known that the capacity of AWGN channel is
	\begin{equation}
		C=\frac{1}{2}\log\left(1+\frac{P}{\sigma_0^2}\right).\label{eqnAWGNcap}
	\end{equation}
	Comparing \eqref{eqnAWGNcap} and \eqref{eqnAWGNlower}, we see that the lower bound is very close to the capacity in the high SNR regime.

	As another example, consider the constraints $X\geq 0$, and $\E{X}\leq \alpha$ on admissible input measures. Here, we obtain the lower bound 
	\begin{equation*}
		\mathop{\max_{\substack{f_W(\cdot):W\geq 0\\\E{W} \leq \frac{\alpha}{\sigma_0}}}{\h{W} -\h{Z}}}
		=\frac{1}{2}\log{\frac{\alpha^2 \ue}{2\pi\sigma_0^2}},
	\end{equation*}
	where we used the fact that the maximum is achieved by the exponential distribution $f_W(w)=\sigma_0/\alpha\exp(-w\sigma_0/\alpha)$ for $w\geq0$ and $f_W(w)=0$ for $w<0$ \cite[Section 12.1]{Cover06}.
	Unlike the first example above, an exact capacity formula for this channel is not known.
\end{example}

%%%%%%%%%%%%%%%
% Lower Bound: d log f + int 1/f
\subsection{Second Idea for Lower Bound} \label{sec:1/f+logCapLB}
Now, we are going to provide another lower bound which is more appropriate in the channels for  which  $Z$ is either non-negative or non-positive, and $\sigma(x)$ is a monotonic function.
An example of such channels is the molecular timing channel discussed in the introduction.

% Theorem: d log f + int 1/f Lower Bound
\begin{theorem} \label{thm:1/f+logCapLB}
	Assume an ASDN channel defined in \eqref{eqn:ChannelModel} with $\sigma:(\ell,u)\mapsto(0,\infty)$ for $-\infty \leq \ell < u \leq +\infty$.
	If $X$ is a continuous random variable with pdf $f_X(x)$, and
	\begin{equation} \label{eqn:1/f+logCapLB:f(x)const}
		\sigma(x) \text{ is continuous and monotonic over } (\ell,u),
	\end{equation}
	\begin{equation} \label{eqn:1/f+logCapLB:1/f(x)const}
		\frac{1}{\sigma(x)} \text{ is Riemann integrable on } (\ell,u),
	\end{equation}
	then 
	\begin{equation*}
		\I{X}{Y}
		\geq \alpha \h{\psi(X)} - \beta,
	\end{equation*}
	provided that $\alpha$, $\beta$ are well-defined, and $\alpha>0$.
	In order to define the variables $\alpha$, $\beta$, and the function $\psi(x)$, take some arbitrary $\delta> 0$ and proceed as follows:
	\begin{itemize}
		\item If the function $\sigma(x)$ is increasing over $(\ell,u)$, let\\
		\begin{equation*}
			\psi(x) = \delta\log \sigma(x) + \int_c^x{\frac{1}{\sigma(t)}\ud t},
		\end{equation*}
\if@twocolumn
		\begin{equation*}
			\alpha = \Pr{Z\geq\delta},
		\end{equation*}
		\begin{equation*}
			\beta = \alpha\h{Z|Z\geq\delta}+\mathrm{H}_2(\alpha),
		\end{equation*}
\else
		\begin{equation*}
			\alpha = \Pr{Z\geq\delta},
			\qquad\beta = \alpha\h{Z|Z\geq\delta}+\mathrm{H}_2(\alpha),
		\end{equation*}
\fi
		\item
		If the function $\sigma(x)$ is decreasing over $(\ell, u)$, let \\
		\begin{equation*} \label{eqn:psi=1/f+logDec}
			\psi(x) = -\delta\log \sigma(x) + \int_c^x{\frac{1}{\sigma(t)}\ud t},
		\end{equation*}
\if@twocolumn
		\begin{equation*}
			\alpha = \Pr{Z\leq\delta},
		\end{equation*}
		\begin{equation*}
			\beta = \alpha\h{Z|Z\leq\delta}+\mathrm{H}_2(\alpha),
		\end{equation*}
\else
		\begin{equation*}
			\alpha = \Pr{Z\leq-\delta},
			\qquad\beta = \alpha\h{Z|Z\leq-\delta}+\mathrm{H}_2(\alpha),
		\end{equation*}
\fi
	\end{itemize}
	where $c\in(\ell,u)$ is arbitrary, and
	\begin{equation*}
		\mathrm{H}_2(p)
		:= -p\log{p}-(1-p)\log{(1-p)}.
	\end{equation*}
\end{theorem}

% Remark
\begin{remark}
	Observe that in both cases, $\psi(x)$ is an strictly increasing function of $x$ defined over $(\ell,u)$, as $\sigma(x)>0$ and $\log(x)$ is increasing.
	Similar to Remark \ref{rmk3}, the choice of $c\in(\ell,u)$ does not affect the value of $\h{\psi(X)}$, and hence the lower bound.
	However, the choice of $\delta>0$ affects the lower bound.  
\end{remark}

The above theorem is proved in Section \ref{subsec:prf:thm:1/f+logCapLB}.

% Corollary: More Constraints
\begin{corollary} \label{cor:1/f+logCapLB:MorConst}
	Similar to Corollary \ref{cor:MajCapLB:MorConst}, let $V=\psi(X)$.
	Since $\psi(\cdot)$ is a one-to-one (strictly increasing) function, we obtain 
	\begin{equation*}
		\max_{\mu_X\in\mathcal{F}\cap \mathcal{AC}}\alpha \h{\psi(X)} - \beta
		=\max_{f_V\in\mathcal{G}}\alpha\h{V} -\beta
	\end{equation*}
	where $\mathcal{F}$ is defined in \eqref{eqn:FeasibleSet fx}, and $V \sim f_V$ belongs to
\if@twocolumn
	\begin{align*}
		\mathcal{G}
		=\{f_V(\cdot)\big| &\mu_V \in \mathcal{AC},\\
		&\textnormal{supp}(\mu_V) \subseteq \psi(\mathcal{X}), \\
		&\mathbb{E}[g_i(\psi^{-1}(V))]\leq 0 \text{ for all } i=1,\cdots, k\}.
	\end{align*}
\else
	\begin{equation*}
		\mathcal{G}
		=\{f_V(\cdot)\big| \mu_V \in \mathcal{AC},
		~\textnormal{supp}(\mu_V) \subseteq \psi(\mathcal{X}),
		~\mathbb{E}[g_i(\psi^{-1}(V))]\leq 0 \text{ for all } i=1,\cdots, k\}.
	\end{equation*}
\fi
	Hence, from Theorem \ref{thm:1/f+logCapLB} we obtain that
	\begin{equation*}
		\max_{\mu_X\in\mathcal{F}}\I{X}{Y}
		\geq \alpha\max_{f_V\in\mathcal{G}}\h{V} -\beta,
	\end{equation*}
	where $\alpha$ and $\beta$ are constants defined in Theorem \ref{thm:1/f+logCapLB}.
\end{corollary}
As mentioned earlier, to maximize $\h{V}$ over $f_V\in\mathcal{G}$, we can use known results on maximum entropy probability distributions, \emph{e.g.,} see \cite[Chapter 3.1]{Ihara93}.

% Corollary: Finite Support
\begin{corollary} \label{cor:1/f+logCapLB:FinSup}
	Consider an ASDN channel satisfying \eqref{eqn:1/f+logCapLB:f(x)const} and \eqref{eqn:1/f+logCapLB:1/f(x)const}.
	Assume that the only input constraint is $\mathcal{X}=(\ell,u)$ \emph{i.e.} $\ell < X < u$.
	Then, from Corollary \ref{cor:1/f+logCapLB:MorConst}, we obtain the lower bound 
\if@twocolumn
	\begin{align*}
		&\alpha\max_{f_V\in\mathcal{G}}{\h{V}} -\beta \\
		&\qquad=\alpha\log
		\left[\delta \left|\log\frac{\sigma(u^-)}{\sigma(\ell^+)}\right|+\int_\ell^u{\frac{1}{\sigma(x)}\ud x}\right]
		-\beta,
	\end{align*}
\else
	\begin{equation*}
		\alpha\max_{f_V\in\mathcal{G}}{\h{V}} -\beta
		=\alpha\log
		\left[\delta \left|\log\frac{\sigma(u^-)}{\sigma(\ell^+)}\right|+\int_\ell^u{\frac{1}{\sigma(x)}\ud x}\right]
		-\beta,
	\end{equation*}
\fi
	where $\alpha$ and $\beta$ are defined in Theorem \ref{thm:1/f+logCapLB}, and
	\begin{equation*}
		\sigma(\ell^+):=\lim_{x\downarrow \ell}{\sigma(x)},
		\qquad \sigma(u^-):=\lim_{x\uparrow u}{\sigma(x)}.
	\end{equation*}
	The lower bound is achieved by taking a uniform distribution for $f_V(w)$ over $\psi(\mathcal{X})$ if this set is bounded \cite[Section 3.1]{Ihara93}.
	Else, if $\psi(\mathcal{X})$ has an infinite length, the capacity is infinity by choosing a pdf $f_V(v)$ such that $\h{V}=+\infty$. (see Example \ref{exp:h=+-inf}).
	The equivalent pdf $f_X(x)$ for $X$ is the pdf of $\psi^{-1}(V)$.
\end{corollary}

%%%%%%%%%%%%%%
% KL Symmetric Upper Bounds
\subsection{An Upper Bound} \label{subsec:KLUpBnd}
We begin by reviewing upper bound given in \cite{moser2012capacity} to motivate our own upper bound.
The upper bound in \cite{moser2012capacity} works by utilizing Topsoe's
inequality \cite{topsoe} to bound mutual information $\I{X}{Y}$ from above as follows:
\begin{equation*}
	\I{X}{Y} \leq \mathbb{E}_{\mu_X}[\D{f(y|x)}{q(y)}].
\end{equation*}
for any arbitrary pdf $q(y)$ on output $Y$.
The distribution $q(y)$ is chosen carefully to allow for calculation of the above KL divergence. The particular form of $\sigma(x)=\sqrt{c_0+c_1x}$ makes explicit calculations possible.
The second difficulty in calculating the above expression is that we need to take expected value over input measure $\mu_X$.
However, the capacity achieving input measure is not known.
This difficulty is addressed by the technique of ``input distributions that escape to infinity", under some assumptions about the peak constraint. 

In this part, we give an upper bound based on the KL symmetrized upper bound of  \cite{aminian2015capacity}. The idea is that
\begin{align*}
	\I{X}{Y}
	&=\mathrm{D}(\mu_{X,Y} \| \mu_X\mu_Y)\\
	&\leq \mathrm{D}(\mu_{X,Y} \| \mu_X\mu_Y)+\mathrm{D}( \mu_X\mu_Y\| \mu_{X,Y})\\
	&\triangleq\mathrm{D}_{\text{sym}}(\mu_{X,Y} \| \mu_X\mu_Y).
\end{align*}
Our upper bound has the advantage of being applicable to a large class of $\sigma(x)$.
To state this upper bound, let $\Cov{X}{Y}:=\E{XY}-\E{X}\E{Y}$ be the covariance function between two random variables $X$ and $Y$. 

% Theorem: KL Symmetric Upper Bound
\begin{theorem} \label{thm:KLUpBnd}
	For any AGSDN channel defined in \eqref{eqn:ChannelModel}, we have
\if@twocolumn
	\begin{align*}
		\I{X}{Y}
		\leq& -\frac{1}{2}\Cov{X^2+\sigma^2(X)}{\frac{1}{\sigma^2(X)}} \\
		&+\Cov{X}{\frac{X}{\sigma^2(X)}},
	\end{align*}
\else
	\begin{equation*}
		\I{X}{Y}
		\leq -\frac{1}{2}\Cov{X^2+\sigma^2(X)}{\frac{1}{\sigma^2(X)}}
		+\Cov{X}{\frac{X}{\sigma^2(X)}},
	\end{equation*}
\fi
	provided that the covariance terms on the right hand side are finite. 
\end{theorem}
The proof can be found in Section \ref{subsec:Proof:thm:KLUpBnd}

% Corollary: KL Capacity Symmetric Upper Bound
\begin{corollary} \label{cor:KLCapUpBnd}
	For an AGSDN channel with parameters $\sigma(x)$, $Z\sim\mathcal{N}(0,1)$, and $\mathcal{X}=[0,u]$, if functions $\sigma(x)$ and ${x}/{\sigma^2(x)}$ are increasing over $\mathcal{X}$, $\sigma(0)>0$, and $x^2+\sigma(x)$ is convex over $\mathcal{X}$ then
	\begin{equation*}
		\max_{\substack{\mu_X:0\leq X \leq u\\\E{X} \leq \alpha}}{\;\I{X}{Y}}
		\leq
		\begin{cases}
			\frac{1}{8} F & \alpha\geq\frac{u}{2}\\
			\frac{1}{2}\left(1-\frac{\alpha}{u}\right)\frac{\alpha}{u}F & \alpha<\frac{u}{2}
		\end{cases},
	\end{equation*}
	where
	\begin{equation*}
		F=\frac{u^2}{\sigma^2(u)}
		+\frac{u^2}{\sigma^2(0)}
		+\frac{\sigma^2(0)}{\sigma^2(u)}
		+\frac{\sigma^2(u)}{\sigma^2(0)}
		-2.
	\end{equation*}
\end{corollary}
The corollary is proved in Section \ref{subsec:Proof:cor:KLCapUpBnd}.

% Remark: C=inf
\begin{remark}
	Even though Corollary \ref{cor:KLCapUpBnd}  is with the assumption $\sigma(0)>0$, if we  formally set $\sigma(0)=0$, we see that $F$ and the upper bound on capacity becomes infinity.
	This is consistent with Theorem \ref{thm:iffC=inf} when $\sigma(0)=0$.
\end{remark}

% Remark: KL Symmetric Upper Bound in Different Applications
\begin{corollary}
	The particular choice of $\sigma(x)=\sqrt{c_0^2+c_1^2x}$ that was motivated by applications discussed in the Introduction has the property that $\sigma(x)$, ${x}/{\sigma^2(x)}$ are increasing and Theorem \ref{thm:KLUpBnd} can be applied.
\end{corollary}

%%%%%%%%
% Useful Lemmas
%%%%%%%%
\section{Some Useful Lemmas} \label{sec:UsefulLemma}
In this section, we provide three lemmas used in the proof of theorems in this paper.

% Lemma: h(Y|X) in ASDN
\begin{lemma} \label{lmm:hY|X}
	In an ASDN channel defined in \eqref{eqn:ChannelModel}, with continuous random variable noise $Z$ with pdf $f_Z(\cdot)$, and noise coefficient $\sigma(x)>0$ $(\mu_X-\text{\emph{a.e.}})$, the conditional measure $\mu_{Y|X}(\cdot|x)$ has the following pdf:
	\begin{equation*}
		f_{Y|X}(y|x) = \frac{1}{\sigma(x)} f_Z\left(\frac{y-x}{\sigma(x)}\right),
		\qquad\mu_{X,Y}\text{-\emph{(a.e.)}}.
	\end{equation*}
	Moreover, $Y$ is a continuous random variable with the pdf 
	\begin{equation*}
		f_Y(y)=\E{\frac{1}{\sigma(X)}f_Z\left(\frac{y-X}{\sigma(X)}\right)}.
	\end{equation*}
	Furthermore, if $\h{Z}$ exists, $\h{Y|X}$ can be defined and is equal to
	\begin{equation*}
		\h{Y|X} = \E{\log \sigma(X)} + \h{Z}.
	\end{equation*}
\end{lemma}
The lemma is proved in Section \ref{subsec:prf:lmm:hY|X}.

% Lemma: h(X)+E[ln f_{\sigma}(x)]
\begin{lemma} \label{lmm:hX+logf=hgX}
	Let $X$ be a continuous random variable with pdf $f_X(x)$.
	For any function $\sigma:(\ell,u)\mapsto [0,+\infty)$ such that $\sigma(x)$ is Riemann integrable over $(\ell,u)$ and $\sigma(x)>0$ (a.e), where $-\infty\leq\ell<u\leq+\infty$, we have that
	\begin{equation} \label{eqn:hX-logf}
		\h{X}+\E{\log{\sigma(X)}}=\h{\varphi(X)}.
	\end{equation}
	where
	\begin{equation} \label{eqn:gX}
		\varphi(x)=\int_c^x{\sigma(t)\ud t},
	\end{equation}
	where $c\in(\ell,u)$ is an arbitrary constant.

	Note that if the left-hand side does not exist, or becomes $\pm\infty$, the same occurs for the right-hand side and \emph{vice versa}.
\end{lemma}
The lemma is proved in Section \ref{subsec:prf:lmm:hX+logf=hgX}.

% Lemma: Using Convex Cover Method
\begin{lemma} \label{lmm:ConvCover}
	Let $X$ be a random variable with probability measure $\mu_X$, and the functions $w(x)$ and $v(x)$ be increasing over $[\ell,u]$, where $-\infty<\ell<u<+\infty$.
	If $v(x)$ is convex over $[\ell,u]$, then
\if@twocolumn
	\begin{align*}
		&\max_{\substack{\mu_X:\ell\leq X \leq u\\\E{X} \leq \alpha}}\Cov{w(X)}{v(X)} \EQnum\label{eqn:OrigProbCov}\\
		&\qquad\leq \beta [w(u)-w(\ell)] [v(u)-v(\ell)],
	\end{align*}
\else
	\begin{equation} \label{eqn:OrigProbCov}
		\max_{\substack{\mu_X:\ell\leq X \leq u\\\E{X} \leq \alpha}}\Cov{w(X)}{v(X)}
		\leq \beta [w(u)-w(\ell)] [v(u)-v(\ell)],
	\end{equation}
\fi
	where
	\begin{align*}
		\beta=
		\begin{cases}
			\frac{1}{4} & \alpha\geq\frac{\ell+u}{2} \\
			\frac{(u-\alpha)(\alpha-\ell)}{(u-\ell)^2} & \alpha<\frac{\ell+u}{2}
		\end{cases}.
	\end{align*}
	Furthermore, for the case $\alpha\geq(\ell+u)/2$, a maximizer  of \eqref{eqn:OrigProbCov} is the pmf
	\begin{equation*}
		p_X(\ell)=p_X(u)=\frac{1}{2}.
	\end{equation*}
	For the case $\alpha<(\ell+u)/2$ if $v(x)$ is linear,  a maximizer  of \eqref{eqn:OrigProbCov} is the pmf
	\begin{equation*}
		p_X(\ell)=1-p_X(u)=\frac{u-\alpha}{u-\ell}.
	\end{equation*}
\end{lemma}
The proof is given in Section \ref{subsec:Proof:lmm:ConvCover}.

%%%%%%%%%%
% Numerical Results
%%%%%%%%%%
\section{Numerical Results}\label{Numericalresults}
\begin{figure*}
	\centering
	\begin{minipage}{0.45\textwidth}
		\includegraphics[trim={3cm 0 0 0},scale=0.4]{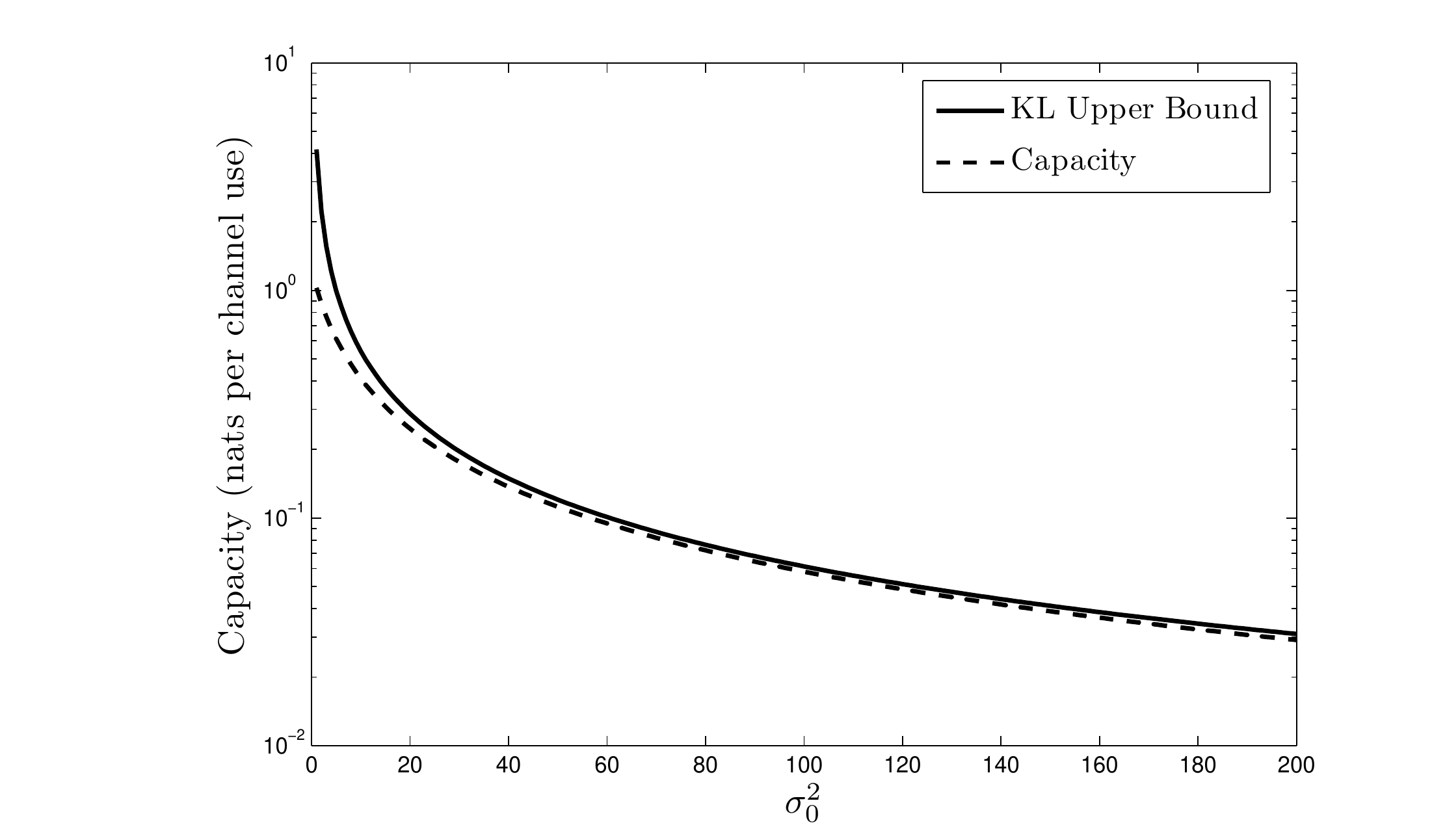}

		\caption{Capacity and Symmetrized divergence upper bound in terms of $c_0^2$ for AGSDN channel with $\mathsf{A}=5$, $\alpha=2.5$, $c_1=1$ and function $\sigma(x)=\sqrt{c_0^2+x}$.} \label{fig2}
	\end{minipage}\qquad
	\begin{minipage}{0.45\textwidth}
		\includegraphics[trim={3cm 0 0 0},scale=0.4]{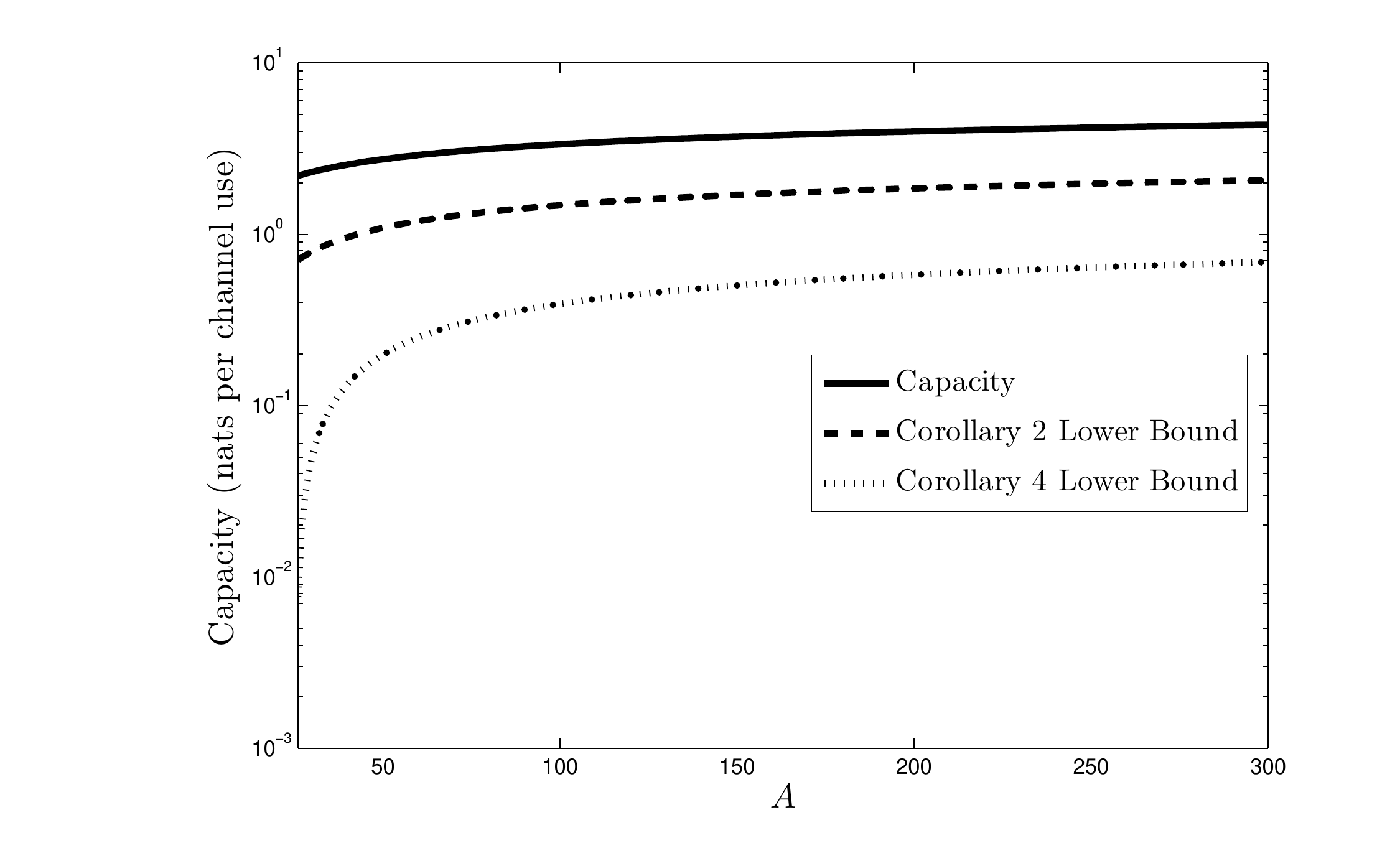}

		\caption{Capacity and lower bound at corollary \ref{cor:MajCapLB:FiniteSupport} and \ref{cor:1/f+logCapLB:FinSup} in terms of $A$ for AGSDN channel with  function $\sigma(x)=\sqrt{1+x}$.} \label{fig3}
	\end{minipage}
\end{figure*}

In this section, some numerical results are given for $\sigma(x)=\sqrt{c_0^2+x}$ and $Z\sim\mathcal{N}(0,1)$. The upper bound, Corollary (\ref{cor:KLCapUpBnd}), and the capacity are depicted in the logarithmic scale in Fig.\ \ref{fig2}, where we have considered the peak constraint $\mathsf{A}=5$ and average constraint $\alpha=2.5$ . It can be observed that the distance between the upper bound and the capacity is a small constant in the logarithmic scale and low SNR regime. This is consistent with \cite{aminian2015capacity} that argues that the upper bound based on symmetrized KL divergence is mostly suitable for the low SNR regime. 

The lower bounds of Corollaries \ref{cor:1/f+logCapLB:FinSup} and \ref{cor:MajCapLB:FiniteSupport} are plotted  in Fig.\ \ref{fig3} for the function $\sigma(x)=\sqrt{c_0^2+x}$ in terms of peak constraint, $A$. Here, $c_0=1$ is assumed. The lower bound of Corollary \ref{cor:MajCapLB:FiniteSupport} for  $0<X<A$ is computed by the following closed form formula:
\if@twocolumn
\begin{align*}
	&\log\left({\int_0^A\frac{1}{\sqrt{c_0^2+x}}}\right)-h(Z) \\
	&\qquad=\log\left(2{\sqrt{A+c_0^2}-2c_0}\right)-\frac{1}{2}\log{2\pi e},
\end{align*}
\else
\begin{equation*}
	\log\left({\int_0^A\frac{1}{\sqrt{c_0^2+x}}}\right)- \h{Z}
	=\log\left(2{\sqrt{A+c_0^2}-2c_0}\right)-\frac{1}{2}\log(2\pi \ue),
\end{equation*}
\fi
while the lower bound of Corollary \ref{cor:1/f+logCapLB:FinSup} equals
\if@twocolumn
\begin{align*}
	&\alpha\log\left(\delta \left|\log\frac{\sigma(A)}{\sigma(0)}\right|+\int_0^A{\frac{1}{\sqrt{c_0^2+x}}\ud x}\right)-\beta\\
	&\qquad=\alpha\log\left(\delta \left|\log\frac{\sqrt{A+c_0^2}}{c_0}\right|+2\sqrt{A+c_0^2}-2c_0\right)-\beta
\end{align*}
\else
\begin{equation*}
	\alpha\log\left(\delta \left|\log\frac{\sigma(A)}{\sigma(0)}\right|+\int_0^A{\frac{1}{\sqrt{c_0^2+x}}\ud x}\right)-\beta
	=\alpha\log\left(\delta \left|\log\frac{\sqrt{A+c_0^2}}{c_0}\right|+2\sqrt{A+c_0^2}-2c_0\right)-\beta
\end{equation*}
\fi
where $\delta >0 $ and
\if@twocolumn
\begin{equation*}
	\alpha = \Pr{Z\geq\delta},
\end{equation*}
\begin{equation*}
	\beta = \alpha\h{Z|Z\geq\delta}-\alpha\log{\alpha}-(1-\alpha)\log{(1-\alpha)}.
\end{equation*}
\else
\begin{equation*}
	\alpha = \Pr{Z\geq\delta},
	\qquad\beta = \alpha\h{Z|Z\geq\delta}-\alpha\log{\alpha}-(1-\alpha)\log{(1-\alpha)}.
\end{equation*}
\fi
We maximized over $\delta$ in order to find the lower bound of Corollary \ref{cor:1/f+logCapLB:FinSup}.
{The first lower bound is better than the second one mainly because of the multiplicative coefficient $\alpha$  of the second lower bound.
Since the second lower bound is for a more general class of channels, we should consider the positive (or negative) part of the support of $Z$, causing a multiplicative of coefficient $1/2$ for the Gaussian noise. However, if the support of $Z$ is positive (or negative) reals, the two lower bounds do not differ much.}

%%%%%
% Proofs
%%%%%
\section{Proofs} \label{sec:Proofs}
%%%%%%%%%%%%%%%%%%%%
% Proof of Theorem: Capacity Achievability
\subsection{Proof of Theorem \ref{thm:CapDef}} \label{subsec:prf:thm:CapDef}
\textbf{Finiteness of capacity:}
	The first step is to show that the capacity is finite:
	\begin{equation} 
		\sup_{\mu_X\in\mathcal{F}}\I{X}{Y} < \infty.
	\end{equation}
	To prove this, it suffices to show that the supremum of both $\h{Y}$ and $\h{Y|X}$ over $\mu_X\in\mathcal{F}$ are finite, \emph{i.e.,}
	\begin{equation} \label{eqn:hY hY|X Bnd}
		|\h{Y}|,|\h{Y|X}|<+\infty, \text{ uniformly on } \mu_X\in\mathcal{F}.
	\end{equation}
	Utilizing Lemma \ref{lmm:hY|X}, the existence and boundedness of $\h{Y|X}$ is obtained as follows:
	\begin{equation*}
		|\h{Y|X}|
		\leq \max\{|\log{\sigma_\ell}| , |\log{\sigma_u}|\}
		+ |\h{Z}|
		< \infty,
	\end{equation*}
	uniformly on $\mathcal{F}$.
	From Lemma \ref{lmm:hY|X}, we obtain that $Y$ is continuous with a pdf $f_Y(y)$.
	To prove that the integral defining $\h{Y}$ is convergent to a finite value (existence of entropy), and furthermore the integral is convergent to a value that is bounded uniformly on $\mathcal F$,
	it is sufficient to show that there are some positive real $\gamma$, $\bar{m}$ and $v$ such that for any $\mu_X\in \mathcal F$, we have \cite{GGA17}:
	\begin{align}
		& \sup_{y\in\bR}{f_Y(y)} < \bar{m}, \label{eqn:fy<a} \\
		& \E{|Y|^\gamma}<v. \label{eqn:E|Y|<a}
	\end{align}
	Also, from Lemma \ref{lmm:hY|X}, we obtain that for any $\mu_X\in\mathcal{F}$
	\begin{equation*}
		f_Y(y) \leq \frac{m}{\sigma_\ell}.
	\end{equation*}
	Thus, \eqref{eqn:fy<a} holds with $\bar{m}={m}/{\sigma_\ell}$.
	In order to prove \eqref{eqn:E|Y|<a}, note that
	\begin{align*}
		\E{|Y|^\gamma}
		\leq & \E{(|X|+\sigma_u|Z|)^\gamma} \\
		\leq & 2^\gamma\E{\max\left\{|X|^\gamma,\sigma_u^\gamma |Z|^\gamma\right\}} \\
		\leq & 2^\gamma\E{|X|^\gamma}
		+ (2\sigma_u)^\gamma \E{|Z|^\gamma} \\
		\leq & 2^\gamma u^\gamma +(2\sigma_u)^\gamma\alpha,
	\end{align*}
	uniformly on $\mathcal F$.
	Thus, $\h{Y}$ is well-defined and uniformly bounded on $\mathcal F$.
	
	Hence, from the definition of mutual information we obtain that
	\begin{equation} \label{eqn:I=h-d-h}
		\I{X}{Y} = \h{Y}-\h{Y|X}
	\end{equation}
	is bounded uniformly for $\mu_X\in\mathcal{F}$.

\textbf{Existence of a maximizer:}
Let
	\begin{equation} \label{eqn:supI<inf}
		C=\sup_{\mu_X\in\mathcal{F}}\I{X}{Y} < \infty.
	\end{equation}
We would like to prove that the above supremum is a maximum. Equation \eqref{eqn:supI<inf} implies existence of a sequence of measures $\{\mu_X^{(k)}\}_{k=1}^\infty$ in $\mathcal F$ such that 
	\begin{equation*}
		\lim_{k\to\infty}{\I{X_k}{Y_k}} = C,
	\end{equation*}
	where $X_k\sim\mu_X^{(k)}$, and $Y_k\sim\mu_Y^{(k)}$ is the output of the channel when the input is $X_k$.
	Furthermore, without loss of generality, we can assume that $\{\mu_X^{(k)}\}_{k=1}^\infty$ is convergent (in the L\'evy measure) to a measure $\mu_X^*\in\mathcal F$.
	The reason is that since $\mathcal{X}$ is compact, the set $\mathcal{F}$ is also compact with respect to the L\'evy measure \cite[Proposition 2]{chan2005capacity}.
	Thus, any sequence of measures in $\mathcal{F}$ has a convergent subsequence.
	With no loss of generality we can take the subsequence as $\{\mu_X^{(k)}\}_{k=1}^\infty$.
	Thus, from convergence in  L\'evy measure, we know that  there is $\mu_X^*\in\mathcal F$ such that 
	\begin{equation} \label{eqn:WeakConv}
		\lim_{k\to\infty}\E{g(X_k)}
		=\E{g(X^*)},
	\end{equation}
	for all $g:\bR\mapsto\mathbb{C}$ such that $\sup_{x\in\bR}{|g(x)|}<+\infty$.
	We would like to prove that
	\begin{equation}\label{eqn:MIC}
		\I{X^*}{Y^*} = C,
	\end{equation}
	where $Y^*\sim\mu_Y^*$ is the output measure of the channel when the input measure is $\mu_X^*$.
	This will complete the proof. 

	From the argument given in the first part of the proof on ``Finiteness of capacity", $\h{Y^*|X^*}$ and $\h{Y^*}$ are well-defined and finite.
	As a result to show \eqref{eqn:MIC}, we only need to prove that
	\begin{equation} \label{eqn:hY|Xn->hY|X*}
		\lim_{k\to\infty}{\h{Y_k|X_k}} = \h{Y^*|X^*},
	\end{equation}
	\begin{equation} \label{eqn:hYn->hY*}
		\lim_{k\to\infty}{\h{Y_k}} = \h{Y^*}.
	\end{equation}
	Since $-\infty<\sigma_\ell<\sigma_u<+\infty$, \eqref{eqn:hY|Xn->hY|X*} is obtained from \eqref{eqn:WeakConv} and Lemma \ref{lmm:hY|X}.

	In order to prove \eqref{eqn:hYn->hY*}, we proceed as follows:
	\begin{itemize}
		\item Step 1: We begin by showing that the sequence $\{\mu_Y^{(k)}\}_{k=1}^\infty$ is a Cauchy sequence with respect to total variation \emph{i.e.}
		\begin{equation} \label{eqn:YnComplete}
			\forall \epsilon>0,\exists N:
			m,n\geq N \Rightarrow \|\mu_Y^{(m)}-\mu_Y^{(n)}\|_V \leq \epsilon,
		\end{equation}
		where for any two arbitrary probability measure $\mu_A$ and $\mu_B$, the total variation distance is defined by \cite[p. 31]{Ihara93}
		\begin{equation*}
			\|\mu_A-\mu_B\|:=
			\sup_{\Delta}{\sum_{i}\big( {\mu_A(E_i)-\mu_B(E_i)}}\big),
		\end{equation*}
		where $\Delta=\{E_1,\cdots,E_m\}\subseteq\mathcal{B}(\bR)$ is the collection of all the available finite partitions.

		\item Step 2: Having established step 1 above, we utilize the fact that the space of probability measures is complete with respect to the total variation metric.
		To show this, note that by Lemma \ref{lmm:hY|X}, all the $Y_k$'s have a pdf, and hence the total variation can be expressed in terms of the $\|\cdot\|_{L_1}$ norm between pdfs \cite[Lemma 1.5.3]{Ihara93}.
		From \cite[p. 276]{Stein05} we obtain that this space of pdfs is complete with respect to $\|\cdot\|_{L_1}$ norm.

		As a result, $\mu_Y^{(k)}$ converges to some measure $\widehat{Y}\sim\widehat{\mu}_Y$ with respect to the total variation metric. We further claim that this convergence implies that 
		\begin{equation} \label{eqn:hYn->hYhat}
			\lim_{k\to\infty}{\h{Y_k}}=\h{\widehat{Y}}.
		\end{equation}
		The reason is that from \eqref{eqn:fy<a} and \eqref{eqn:E|Y|<a}, we see that $\{f_Y^{(k)}\}$ and $f_{\hat{Y}}$ are uniformly bounded and have finite $\gamma$-moments.
		Therefore, \eqref{eqn:hYn->hYhat} follows from \cite[Theorem 1]{GGA17}.
		Thus, in step 2, we obtain that the sequence $\h{Y_k}$ has a limit.

		\item Step 3: We show that the limit found in Step 2 is equal to $\h{Y^*}$, \emph{i.e.,}
		\begin{equation} \label{eqn:hYhat=hY*}
			\h{\widehat{Y}}=\h{Y^*}.
		\end{equation}
		This completes the proof of \eqref{eqn:hYn->hY*}.
	\end{itemize}

	Hence, it only remains to prove \eqref{eqn:YnComplete} and \eqref{eqn:hYhat=hY*}.
	
% Yn is Cauchy
	\emph{Proof of \eqref{eqn:YnComplete}}: 
	Since $\{\I{X_k}{Y_k} \}_{k=1}^\infty$ is convergent to $C$, for any $\epsilon'>0$, there exists $N$ such that:
	\begin{equation*}
		|C-\I{X_k}{Y_k} |\leq\epsilon',
		\qquad \forall k\geq N.
	\end{equation*}
	Now, consider $m,n\geq N$.
	Let $Q$ be a uniform Bernoulli random variable, independent of all previously defined variables.
	When $Q=0$, we sample from measure $\mu_X^{(m)}$ and when $Q=1$, we sample from measure $\mu_X^{(n)}$.
	This induces the measure  $\widetilde{X}\sim\mu_{\widetilde{X}}$ defined as follows:
	\begin{equation*}
		\mu_{\widetilde{X}} = \frac{1}{2}\mu_X^{(m)} + \frac{1}{2} \mu_X^{(n)}.
	\end{equation*}
	Let $\widetilde{Y}\sim\mu_{\widetilde{Y}}$ be the output of the channel when the input is $\widetilde{X}$.
	We have a Markov chain $Q-\widetilde{X}-\widetilde{Y}$. Note that
	\begin{equation*}
		\I{\widetilde{X}}{\widetilde{Y}|Q}=\frac{1}{2} \I{X_m}{Y_m} + \frac{1}{2}\I{X_n}{Y_n}.
	\end{equation*}
	From concavity of mutual information in input measure, we obtain that:
\if@twocolumn
	\begin{align*}
		\I{\widetilde{X}}{\widetilde{Y}}
		&\geq \frac{1}{2} \I{X_m}{Y_m} + \frac{1}{2}\I{X_n}{Y_n} \\
		&\geq  C-\epsilon'.
	\end{align*}
\else
	\begin{equation*}
		\I{\widetilde{X}}{\widetilde{Y}}
		\geq \frac{1}{2} \I{X_m}{Y_m} + \frac{1}{2}\I{X_n}{Y_n}
		\geq  C-\epsilon'.
	\end{equation*}
\fi
	Since $\mathcal{F}$ is an intersection of half spaces, it is convex and as a result $\mu_{\widetilde{X}}\in\mathcal{F}$.
	Thus, $\I{\widetilde{X}}{\widetilde{Y}}\leq C$, and we obtain that
	\begin{equation*}
		\I{\widetilde{X}}{\widetilde{Y}} - \I{\widetilde{X}}{\widetilde{Y}\big|Q}
		\leq \epsilon'.
	\end{equation*}
	Because of the Markov chain $Q - \widetilde{X} - \widetilde{Y}$, we obtain $\I{\widetilde{Y}}{Q\big|\widetilde{X}}=0$ and as a result:
	\begin{equation*}
		\I{\widetilde{Y}}{Q} \leq \epsilon'
		\Longrightarrow \D{\mu_{\widetilde{Y},Q}}{\mu_{\widetilde{Y}}\mu_Q} \leq \epsilon'.
	\end{equation*}
	From the Pinsker's inequality we obtain that
	\begin{equation} \label{eqn:|YQ-YQ|<e}
		\|\mu_{\widetilde{Y},Q}-\mu_{\widetilde{Y}}\mu_Q\|_V \leq \sqrt{2\epsilon'},
	\end{equation}
	where $\|\mu_{\widetilde{Y},Q}-\mu_{\widetilde{Y}}\mu_Q\|_V$ is the total variation between the measures $\mu_{\widetilde{Y},Q}$ and $\mu_{\widetilde{Y}}\mu_Q$.
Note that
	\begin{equation} \label{eqn:|YQ-YQ|=||}
		\|\mu_{\widetilde{Y},Q}-\mu_{\widetilde{Y}}\mu_Q\|_V
		=\frac{1}{2}\|\mu_Y^{(m)}-\mu_{\widetilde{Y}}\|_V
		+\frac{1}{2}\|\mu_Y^{(n)}-\mu_{\widetilde{Y}}\|_V.
	\end{equation}
	Therefore from \eqref{eqn:|YQ-YQ|<e} and \eqref{eqn:|YQ-YQ|=||}, we obtain that
	\begin{equation*}
		\|\mu_Y^{(m)}-\mu_{\widetilde{Y}}\|_V,\|\mu_Y^{(n)}-\mu_{\widetilde{Y}}\|_V\leq 2\sqrt{2\epsilon'}.
	\end{equation*}
	As a result,
	\begin{equation*}
		\|\mu_Y^{(m)}-\mu_Y^{(n)}\|_V \leq 4 \sqrt{2\epsilon'}.
	\end{equation*}
	Hence, by taking $\epsilon'\leq\epsilon^2/32$, we obtain that $\{\mu_Y^{(k)}\}_{k=1}^\infty$ is a Cauchy sequence.
	
%Yhat=Y*
	\emph{Proof of \eqref{eqn:hYhat=hY*}}:
	To this end, it suffices to prove that
	\begin{equation*}
		\Phi_{\widehat{Y}}(\omega)=\Phi_{Y^*}(\omega),
		\qquad \forall \omega\in\bR,
	\end{equation*}
	where $\Phi_X(\omega):=\E{\exp(\uj \omega X)}$ is the characteristic function of the random variable $X$.
	
	Since $Y_k$ converge to $\widehat{Y}$ in total variation, and the fact that convergence in total variation is stronger than weakly convergence \cite[p. 31]{Ihara93}, from \eqref{eqn:WeakConv} we obtain that their characteristic functions, $\Phi_{Y_k}(\omega)$, also converge to $\Phi_{\widehat{Y}}(\omega)$ pointwise.
	
	Hence, it suffices to prove that $\Phi_{Y_k}(\omega)$ converge to $\Phi_{Y^*}(\omega)$ pointwise.
	From \eqref{eqn:ChannelModel}, we obtain that
\if@twocolumn
	\begin{align*}
		\Phi_{Y_k}(\omega)
		= & \E{\ue^{\uj\omega (X_k+\sigma(X_k) Z)}} \\
		= & \E{\ue^{\uj\omega X_k} \Phi_{Z}(\sigma(X_k)\omega)}.
	\end{align*}
\else
	\begin{equation*}
		\Phi_{Y_k}(\omega)
		= \E{\ue^{\uj\omega (X_k+\sigma(X_k) Z)}}
		= \E{\ue^{\uj\omega X_k} \Phi_{Z}(\sigma(X_k)\omega)}.
	\end{equation*}
\fi
	Similarly,
	\begin{equation*}
		\Phi_{Y^*}(\omega)
		= \E{\ue^{\uj\omega X^*} \Phi_{Z}(\sigma(X^*)\omega)}.
	\end{equation*}	
	Since $\{X_k\}$ converges to $X^*$ in L\'evy measure and the function $g(x)=\ue^{\uj\omega x} \Phi_{Z}(\sigma(x)\omega)$ is bounded:
	\begin{equation*}
		|g(x)|=\left|\ue^{\uj\omega x} \Phi_{Z}(\sigma(x)\omega)\right|
		\leq \left|\ue^{\uj\omega x}\right| \left|\Phi_{Z}(\sigma(x)\omega)\right|
		\leq 1,
	\end{equation*}
	from \eqref{eqn:WeakConv} we obtain that $\mathbb{E}[g(X_k)]=\Phi_{Y_k}(\omega)$ converges to $\mathbb{E}[g(X^*)]=\Phi_{Y^*}(\omega)$ pointwise.
	
	{\textbf{Uniqueness of the output pdf:}
	The proof is the same as the first part of the proof of \cite[Theorem 1]{chan2005capacity}.

	This completes the proof.
\qed

%%%%%%%%%%%%%%%%%%%%%%%%%
% Proof of Theorem: Sufficient Condition for I(X;Y)=inf
\subsection{Proof of Theorem \ref{thm:iffC=inf}} \label{subsec:prf:thm:iffC=inf}
	For a continuous input measure, we utilize a later result in the paper, namely Theorem \ref{thm:1/f+logCapLB} by choosing $\ell=c$, $u=c'$ when $c'>c$, or $\ell=c'$, $u=c$ when $c'<c$.
	To use Corollary \ref{cor:1/f+logCapLB:FinSup}, observe that the image of $\mathcal{E}$  under $\psi(\cdot)$ has infinite length.
	This is because the sequence $\{\tilde{x}_i\}$ in $\mathcal E$ was such that the monotone function $\sigma(\cdot)$ converged to zero or infinity on that sequence.
	Then, it is obtained that any pdf $f_X(\cdot)$ such that $\h{\psi(X)}=+\infty$, makes $\I{X}{Y}$ infinity if $|\h{Z|Z>\delta}|<\infty$ (which leads to $|\beta|<\infty$), where $\psi(x)$ is the bijective function of $x$ defined in the statement of Theorem \ref{thm:1/f+logCapLB}.

	In order to prove that $|\h{Z|Z>\delta}|<\infty$, let the random variable $\bar{Z}$ be $Z$ conditioned to $Z>\delta$.
	Due to the continuity of $Z$ and the fact that $\Pr{Z>\delta}>0$, we obtain that $\bar{Z}$ has a valid pdf $f_{\bar{Z}}(z)$ defined by
	\begin{equation*}
		f_{\bar{Z}}(z) =
		\begin{cases}
			\frac{1}{\theta} f_{Z}(z) & z>\delta \\
			0 & z\leq\delta
		\end{cases},
	\end{equation*}
	where $\theta:=\Pr{Z>\delta}>0$.
	Since $\h{Z}$ exists and $|\h{Z}|<\infty$, we obtain that
	\begin{equation*}
		\E{\left|\frac{1}{f_{Z}(Z)}\right|}<\infty.
	\end{equation*}
	Hence,
\if@twocolumn
	\begin{align*}
		|\h{\bar{Z}}| \leq & \E{\left|\frac{1}{f_{\bar{Z}}(\bar{Z})}\right|} \\
		\leq & -\log\theta + \frac{1}{\theta}\E{\left|\frac{1}{f_{Z}(Z)}\right|}
		<\infty.
	\end{align*}
\else
	\begin{equation*}
		|\h{\bar{Z}}|
		\leq \E{\left|\frac{1}{f_{\bar{Z}}(\bar{Z})}\right|}
		\leq -\log\theta + \frac{1}{\theta}\E{\left|\frac{1}{f_{Z}(Z)}\right|}
		<\infty.
	\end{equation*}
\fi
	Therefore, $\h{Y|Z>\delta}$ exists and $|\h{Y|Z>\delta}|<\infty$.
	A similar treatment can be used to prove $|\h{Y|Z<\delta}|<\infty$.

	It remains to construct a discrete pmf with infinite mutual information. The statement of the theorem assumes existence of a sequence $\{\tilde{x}_i\}$ in an open interval $\mathcal{E}=(c, c')\subset\mathcal X$ (or $\mathcal{E}=(c',c)$ if $c'<c$) such that \begin{enumerate}
\item $c$ is the limit of the sequence $\{\tilde{x}_i\}$, 
\item $\sigma(\tilde x_i)$ converges to $0$ or $+\infty$
\item $\sigma(\cdot)$ is monotone and continuous over $\mathcal{E}$
\end{enumerate}
We now make the following claim about existence of another sequence $\{x_i\}_{i=1}^\infty\subseteq\mathcal{E}$ with certain nice properties:

	\textbf{Claim:} Suppose that one cannot find a non-empty interval $[x', x'']\subset\mathcal{E}$ such that $\sigma(x)=0$ for all $x\in[x',x'']$.  Then, there exists $0<a<b$ and a sequence $\{x_i\}_{i=1}^\infty\subseteq\mathcal{E}$, such that
	\begin{itemize}
		\item If $\sigma(x)$ is increasing,
\if@twocolumn
		\begin{align}
			&\Pr{a<Z<b} > 0, \label{eqn:sigmainc1}\\
			&[x_i+a \sigma(x_i) , x_i+b \sigma(x_i)]\cap[x_j+a \sigma(x_j) , x_j+b \sigma(x_j)]\nonumber \\
			&\qquad=\varnothing,
			\qquad\forall i\neq j\in\mathbb{N}.\label{eqn:sigmainc2}
		\end{align}
\else
		\begin{align}
			&\Pr{a<Z<b} > 0, \label{eqn:sigmainc1}\\
			&(x_i+a \sigma(x_i) , x_i+b \sigma(x_i))\cap(x_j+a \sigma(x_j) , x_j+b \sigma(x_j))
			=\varnothing,
			\qquad \forall i\neq j\in\mathbb{N} \label{eqn:sigmainc2} \\
			& 0<\sigma(x_i)<\infty,\qquad \forall i\in\mathbb{N}. \label{eqn:sigmainc3}
		\end{align}
\fi
		\item If $\sigma(x)$ is decreasing,
\if@twocolumn
		\begin{align*}
			&\Pr{-b<Z<-a} > 0, \\
			&(x_i-b \sigma(x_i) , x_i-a \sigma(x_i))\cap(x_j-b \sigma(x_j) , x_j-a \sigma(x_j)) \\
			&\qquad=\varnothing,
			\qquad\forall i\neq j\in\mathbb{N}.
		\end{align*}
\else
		\begin{align*}
			&\Pr{-b<Z<-a} > 0, \\
			&(x_i-b \sigma(x_i) , x_i-a \sigma(x_i))\cap(x_j-b \sigma(x_j) , x_j-a \sigma(x_j))
			=\varnothing,
			\qquad\forall i\neq j\in\mathbb{N} \\
			& 0<\sigma(x_i)<\infty,\qquad \forall i\in\mathbb{N}.
		\end{align*}
\fi
	\end{itemize}
	
	We continue with the proof assuming that this claim is correct; we give the proof of this claim later.
	{To show how this claim can be used to construct a discrete pmf with infinite mutual information, consider the possibility that the assumption of the claim fails: $\sigma(x)=0$ for all $x \in [x',x'']$, then $Y=X$ in that interval when $X\in [x',x'']$. 
	Therefore, we can provide any discrete distribution in that interval such that $\H{X}=\infty$, as a result $\I{X}{Y}=\I{X}{X}=\H{X}=\infty$.} 

	Thus, we should only consider the case that the assumption of the claim holds. Assume that  $\sigma(x)$ is increasing.
	The construction when $\sigma(x)$ is decreasing  is similar.
	Fix a given $a$, $b$, $\{x_i\}_{i=1}^\infty$ satisfying \eqref{eqn:sigmainc1} and  \eqref{eqn:sigmainc2}.
	Take an arbitrary pmf $\{p_i\}_{i=1}^\infty$ such that
	\begin{equation}\label{eqn:pseq}
		\sum_i{p_i \log\frac{1}{p_i}}=+\infty.
	\end{equation}
	Then, we define a discrete random variable $X$, taking values in $\{x_i\}_{i=1}^\infty$ such that $\Pr{X=x_i}=p_i$.
	We claim that $\I{X}{Y} = +\infty$.
	To this end, it suffices to show
\if@twocolumn
	\begin{align}
		\I{X}{Y} \geq & \Pr{a<Z<b}\I{X}{Y|a<Z<b} \nonumber\\
		&- \mathrm{H}_2(\Pr{a<Z<b}), \label{eqn:prf:thm:iffC=inf:I>I|Z>d}
	\end{align}
\else
	\begin{equation} \label{eqn:prf:thm:iffC=inf:I>I|Z>d}
		\I{X}{Y} \geq \Pr{a<Z<b}\I{X}{Y|a<Z<b}
		- \mathrm{H}_2(\Pr{a<Z<b}),
	\end{equation}
\fi
	\begin{equation} \label{eqn:prf:thm:iffC=inf:I|Z>d=inf}
		\I{X}{Y|a<Z<b}=\infty.
	\end{equation}

	\emph{Proof of \eqref{eqn:prf:thm:iffC=inf:I>I|Z>d}}:
	Define random variable $E$ as following:
	\begin{equation*}
		E=
		\begin{cases}
			0&Z\in (a,b)\\
			1&Z\notin (a,b)
		\end{cases}
	\end{equation*}
	From the definition of mutual information, we have that
\if@twocolumn
	\begin{align*}
		\I{X}{Y|E}-\I{X}{Y}
		=&\I{Y}{E|X}-\I{Y}{E}\\
		\leq&\H{E},
	\end{align*}
\else
	\begin{equation*}
		\I{X}{Y|E}-\I{X}{Y}
		=\I{Y}{E|X}-\I{Y}{E}
		\leq \H{E},
	\end{equation*}
\fi
	Since
\if@twocolumn
	\begin{align*}
		\I{X}{Y|E}
		=&\Pr{E=0}\I{X}{Y|E=0}\\
		&+\Pr{E=1}\I{X}{Y|E=1},
	\end{align*}
\else
	\begin{equation*}
		\I{X}{Y|E}
		=\Pr{E=0}\I{X}{Y|E=0}
		+\Pr{E=1}\I{X}{Y|E=1},
	\end{equation*}
\fi
	we conclude \eqref{eqn:prf:thm:iffC=inf:I>I|Z>d}.
	
	\emph{Proof of \eqref{eqn:prf:thm:iffC=inf:I|Z>d=inf}}:
	Since
	\begin{equation*}
		\I{X}{Y|a<Z<b}
		=\H{X}-\H{X|Y,a<Z<b},
	\end{equation*}
	it suffices to show that 
	\begin{equation} \label{eqn:prf:thm:iffC=inf:HX=inf HX|Y=0}
		\H{X} = \infty,
		\qquad \H{X|Y,a<Z<b} = 0.
	\end{equation}
	The  equality $\H{X}:=-\sum{p_i\log p_i}=+\infty$ follows \eqref{eqn:pseq}. 
	To prove the other equality, note that $Y$ belongs to the interval $(x_i+a \sigma(x_i), x_i+b \sigma(x_i))$ when $X=x_i$.
	Therefore, since the intervals $(x_i+a \sigma(x_i), x_i+b \sigma(x_i))$ are disjoint, $X$ can be found from $Y$.
	Thus, $X$ is a function of $Y$ when $a<Z<b$.
	As a result, the second equality of \eqref{eqn:prf:thm:iffC=inf:HX=inf HX|Y=0} is proved. 

	Now, it only remains to prove our Claim on the existence of $a$, $b$, and $\{x_i\}_{i=1}^\infty$. 
	
	We assume that  $\sigma(x)$ is increasing.
	The proof when $\sigma(x)$ is decreasing is similar.
	From the assumptions on $Z$ that $\Pr{Z\geq\delta}>0$, we obtain there exists $\delta<b<\infty$ such that $\Pr{\delta<Z<b}>0$.
	As a result, we select $a=\delta$.

Since $\sigma(x)$ is monotone, we cannot have $\sigma(x')=\sigma(x'')=0$ for two arbitrary distinct $x'$ and $x''$ in $\mathcal{E}$ since this implies that  $\sigma(x)=0$ for all $x$ in between $x'$ and $x''$. As a result, we shall not worry about the constraint \eqref{eqn:sigmainc3} on $\{x_i\}$ because $\sigma(x_i)=0$ can occur for at most one index $i$ and we can delete that element from the sequence to ensure \eqref{eqn:sigmainc3}.

	To show the existence of $\{x_i\}_{i=1}^\infty$, we provide a method to find $x_{i+1}$ with respect to $x_i$.
The method is described below and illustrated in Figure \ref{fig:prf:thm:CInf}.
	% Figures
	\begin{figure*}
\if@twocolumn
		\centering
		\begin{minipage}{0.45\textwidth}
			\includegraphics[scale=0.3]{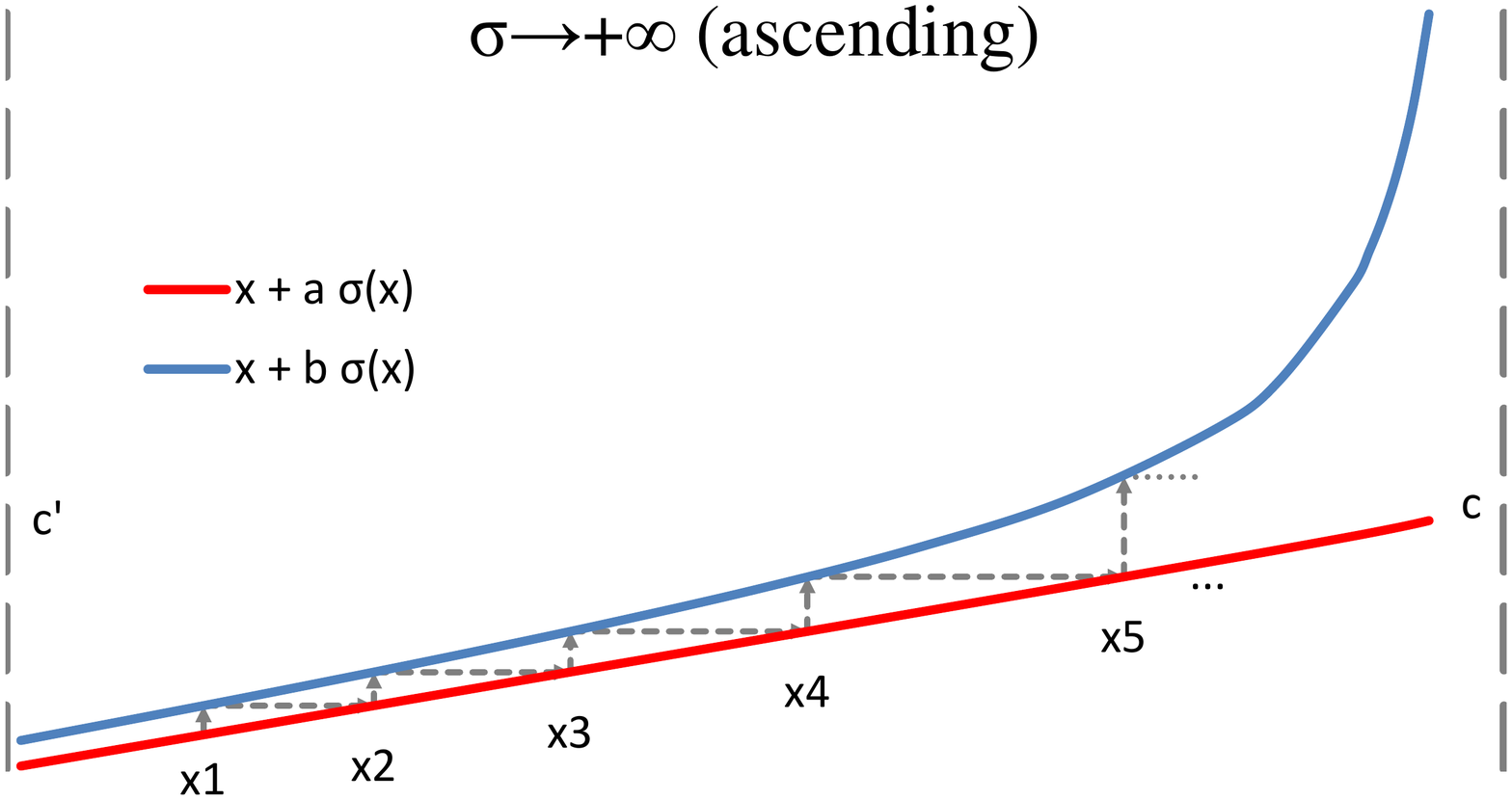}
		\end{minipage}

		\begin{minipage}{0.45\textwidth}
			\includegraphics[scale=0.3]{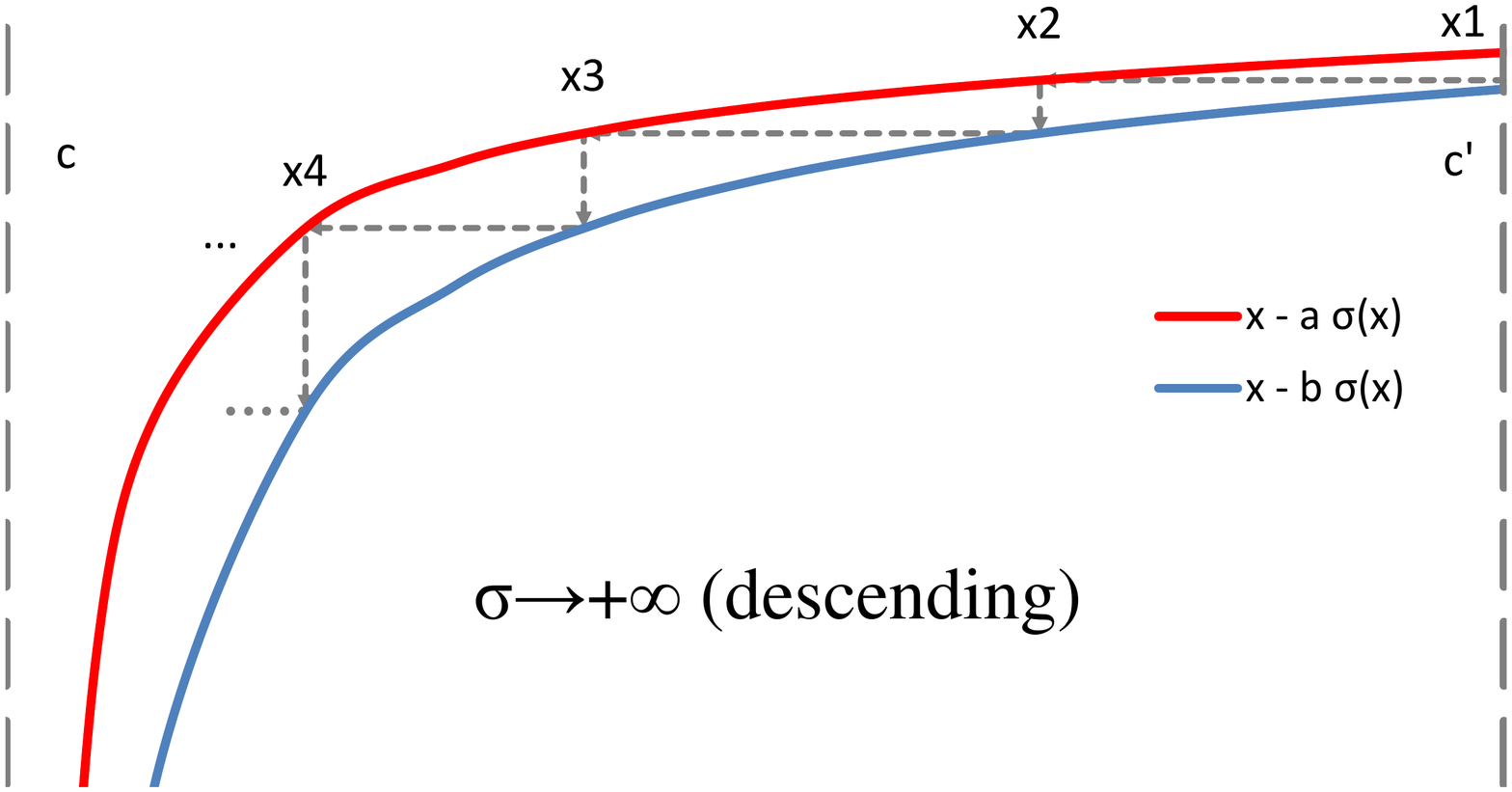}
		\end{minipage}

		\begin{minipage}{0.45\textwidth}
			\includegraphics[scale=0.3]{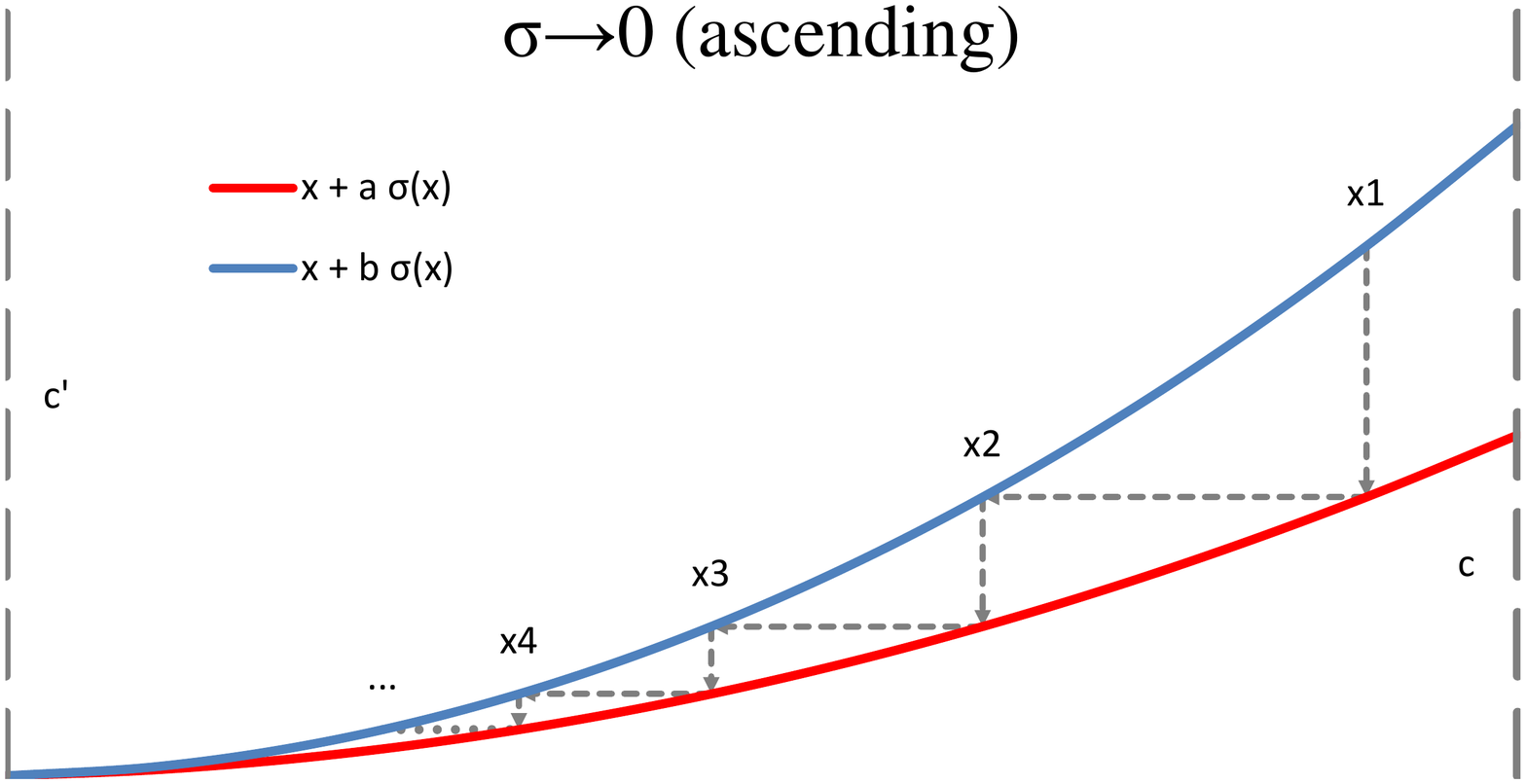}
		\end{minipage}

		\begin{minipage}{0.45\textwidth}
			\includegraphics[scale=0.3]{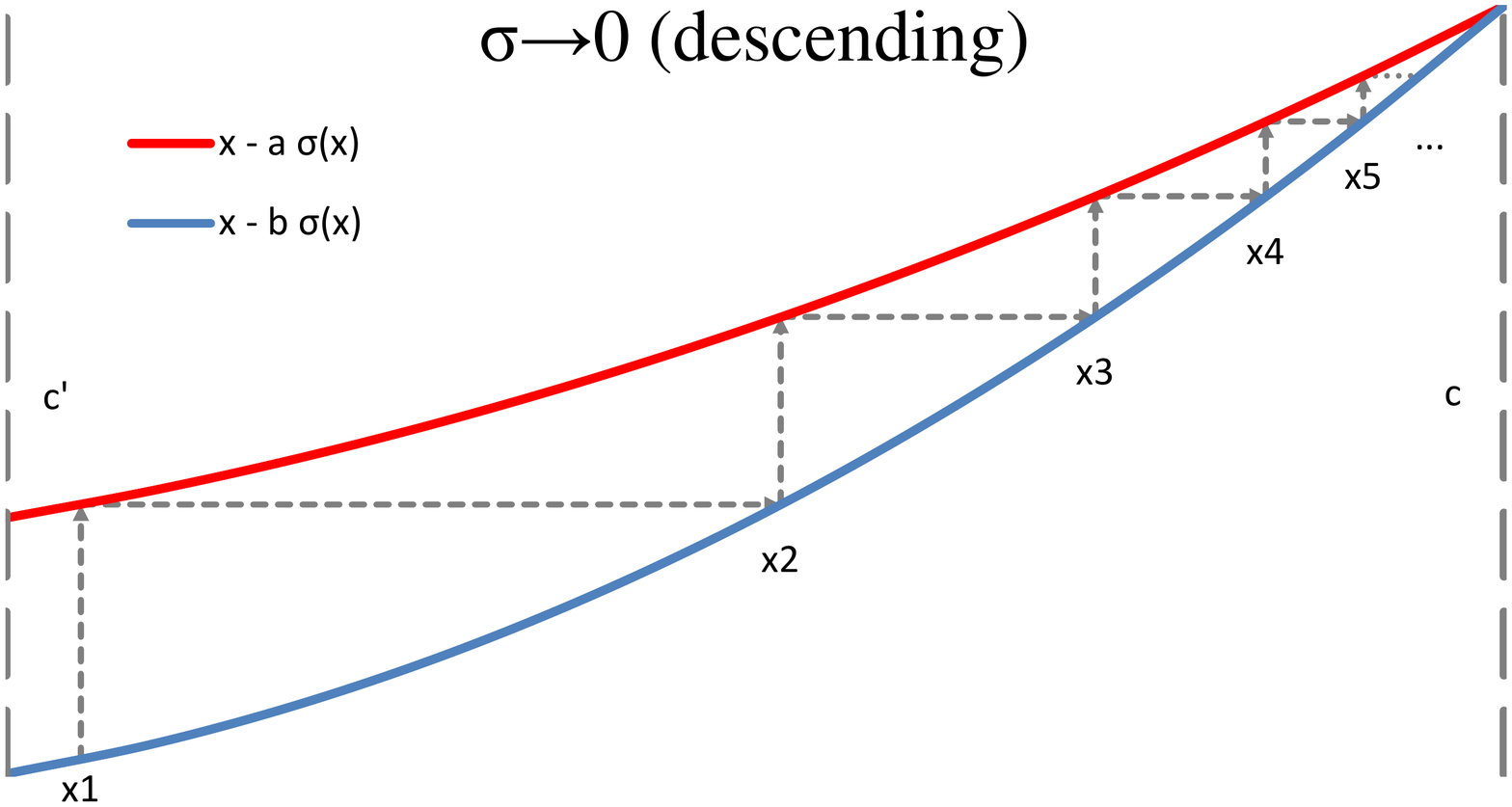}
		\end{minipage}
\else
		\begin{minipage}{0.45\textwidth}
			\includegraphics[scale=0.3]{prf_inf1}
		\end{minipage} \qquad
		\begin{minipage}{0.45\textwidth}
			\includegraphics[scale=0.3]{prf_inf2}
		\end{minipage}

		\begin{minipage}{0.45\textwidth}
			\includegraphics[scale=0.3]{prf_inf4}
		\end{minipage} \qquad
		\begin{minipage}{0.45\textwidth}
			\includegraphics[scale=0.3]{prf_inf3}
		\end{minipage}
\fi
		\caption{Possible cases for $\sigma(x)$ when $|c|<\infty$.} \label{fig:prf:thm:CInf}
	\end{figure*}

	Take $x_1$ an arbitrary element of $\mathcal{E}$.
	Observe that since $\sigma(x)$ is continuous and increasing over $\mathcal{E}$, the functions $x + a \sigma(x)$ and $x + b \sigma(x)$ are continuous and strictly increasing over $\mathcal{E}$, as well as
	\begin{equation*}
		x + a \sigma(x) < x + b \sigma(x),
		\qquad \forall x\in\mathcal{E}.
	\end{equation*}
	Therefore, for the case $c'>c$ (happening when $\sigma(x_i)$ converge to $0$), 
	\begin{equation*}
		\lim_{x\to c}{x+a\sigma(x)}
		=\lim_{x\to c}{x+b\sigma(x)}
		=c,
	\end{equation*}
	Hence, for a given $x_i\in\mathcal{E}$, due to the intermediate value theorem, there exists unique $x_{i+1}$ satisfying $c < x_{i+1}< x_i < c'$ such that
	\begin{equation*}
		x_{i+1} + b \sigma(x_{i+1})
		= x_i + a \sigma(x_i).
	\end{equation*}
	Similarly, for the case $c'<c$ (happening when $\sigma(\tilde x_i)$ converge to $+\infty$), if $x_i \in\mathcal{E}$, there exists unique $x_{i+1}$ satisfying $c > x_{i+1} > x_i>c'$ such that
	\begin{equation*}
		x_{i+1} + a \sigma(x_{i+1})
		= x_i + b \sigma(x_i).
	\end{equation*}
	It can be easily obtained that the intervals created this way are disjoint, and the process will not stop after finite steps.
	Therefore, the theorem is proved.
\qed

%%%%%%%%%%%%%%%%%%%%%%%%%%
% Proof of Theorem: Majorization Capacity Lower Bound
\subsection{Proof of Theorem \ref{thm:MajCapLB}} \label{subsec:Proof:thm:MajCapLB}
	From Lemma \ref{lmm:hY|X} we obtain that $\h{Y|X}$ exists.
	Hence, utilizing Lemma \ref{thm:Maj hY>hX} , we can write
	\begin{equation} \label{eqn:prf:thm:MajCapLB:I1}
		\h{Y} \geq \h{X}
		\Longrightarrow \I{X}{Y}
		\geq \h{X}-\h{Y|X},
	\end{equation}
	provided that
	\begin{equation*}
		\int_\ell^u{f_{Y|X}(y|x)\ud x}
		\leq 1,
	\end{equation*}
	where it is satisfied due to
	\begin{equation*}
		\int_\ell^u{f_{Y|X}(y|x)\ud x}
		= \int_\ell^u{\frac{1}{\sigma(x)} f_Z\left(\frac{y-x}{\sigma(x)}\right) \ud x}
		\leq 1,
	\end{equation*}
	where the last inequality comes from the assumption of the theorem.
	From Lemma \ref{lmm:hY|X}, we have that
	\begin{equation*}
		\h{Y|X} = \E{\log \sigma(X)} + \h{Z}.
	\end{equation*}
	Therefore, \eqref{eqn:prf:thm:MajCapLB:I1} can be written as 
	\begin{equation*}
		\I{X}{Y}
		\geq \h{X}-\E{\log \sigma(X)}-\h{Z}.
	\end{equation*}
	Exploiting Lemma \ref{lmm:hX+logf=hgX} we obtain that
	\begin{equation*}
		\h{X}-\E{\log \sigma(X)}
		=\h{\varphi(X)},
	\end{equation*}
	where $\varphi(X)$ is defined in \eqref{eqn:phi=1/f}
	Hence, the proof is complete.
\qed

%%%%%%%%%%%%%%%%%%%%%%%
% Proof of Theorem: d log f + int 1/f Lower Bound
\subsection{Proof of Theorem \ref{thm:1/f+logCapLB}} \label{subsec:prf:thm:1/f+logCapLB}
	We only prove the case that $\sigma(x)$ is an increasing function over $(\ell, u)$.
	The proof of the theorem for decreasing functions is similar to the increasing case and we only need to substitute $Z\geq\delta$ with $Z\leq -\delta$.
	We claim that
	\begin{equation} \label{eqn:prf:thm:1/f+logCapLB:I>I|Z>d}
		\I{X}{Y} \geq \alpha \I{X}{Y|Z\geq\delta} - \mathrm{H}_2(\alpha).
	\end{equation}
	Consider random variable $E$ as following:
	\begin{equation*}
		E=
		\begin{cases}
			0&Z\geq\delta\\
			1&Z<\delta
		\end{cases}.
	\end{equation*}
	From the definition of mutual information, we have that
\if@twocolumn
	\begin{align*}
		\I{X}{Y|E}-\I{X}{Y}
		=&\I{Y}{E|X}-\I{Y}{E}\\
		\leq&\H{E},
	\end{align*}
\else
	\begin{equation*}
		\I{X}{Y|E}-\I{X}{Y}
		=\I{Y}{E|X}-\I{Y}{E}
		\leq \H{E},
	\end{equation*}
\fi
	Therefore, since
\if@twocolumn
	\begin{align*}
		\I{X}{Y|E}
		=&\Pr{Z\geq\delta}\I{X}{Y|Z\geq\delta}\\
		&+\Pr{Z<\delta}\I{X}{Y|Z<\delta},
	\end{align*}
\else
	\begin{equation*}
		\I{X}{Y|E}
		=\Pr{Z\geq\delta}\I{X}{Y|Z\geq\delta}
		+\Pr{Z<\delta}\I{X}{Y|Z<\delta},
	\end{equation*}
\fi
	we conclude \eqref{eqn:prf:thm:1/f+logCapLB:I>I|Z>d}.

	Now, we find a lower bound for $\I{X}{Y|Z\geq\delta}$.
	From Lemma \ref{lmm:hY|X} we obtain that $Y$ is a continuous random variable.
	We claim that
\if@twocolumn
	\begin{align}
		\I{X}{Y|Z\geq\delta}
		=&\h{Y|Z\geq\delta}-\h{Y|X,Z\geq\delta} \nonumber\\
		=&\h{Y|Z\geq\delta} \nonumber\\
		&- \left( \E{\log{\sigma(X)}} + \h{Z|Z\geq\delta} \right) \label{eqn:prf:thm:1/f+logCapLB:hY|X}\\
		=&\h{Y|Z\geq\delta}-\h{X} \nonumber\\
		&-\E{\log(1+Z\sigma'(X))|Z\geq\delta} \nonumber\\
		&+\h{X} + \E{\log\frac{1+Z\sigma'(X)}{\sigma(X)}\Big|Z\geq\delta} \nonumber\\
		&-\h{Z|Z\geq\delta}, \label{eqn:IXY|Z}
	\end{align}
\else
	\begin{align}
		\I{X}{Y|Z\geq\delta}=&\h{Y|Z\geq\delta}-\h{Y|X,Z\geq\delta} \nonumber\\
		=&\h{Y|Z\geq\delta}
		- \left( \E{\log{\sigma(X)}} + \h{Z|Z\geq\delta} \right) \label{eqn:prf:thm:1/f+logCapLB:hY|X}\\
		=&\h{Y|Z\geq\delta}-\h{X}
		-\E{\log(1+Z\sigma'(X))|Z\geq\delta} \nonumber\\
		&+\h{X} + \E{\log\frac{1+Z\sigma'(X)}{\sigma(X)}\Big|Z\geq\delta}
		-\h{Z|Z\geq\delta}, \label{eqn:IXY|Z}
	\end{align}
\fi
	where  \eqref{eqn:prf:thm:1/f+logCapLB:hY|X} is obtained from Lemma \ref{lmm:hY|X}, and the fact that random variable $Z$ conditioned to $Z\geq\delta$ is also continuous when $\Pr{Z\geq\delta}>0$.
	Moreover, \eqref{eqn:IXY|Z} is obtained by adding and subtracting the term $\E{\log(1+Z\sigma'(X))|Z\geq\delta}$.
	Note that we had not assumed that $\sigma(x)$ needs to be differentiable.
	We had only assumed that $\sigma:(\ell,u)\mapsto(0,\infty)$ is continuous and monotonic over $(\ell,u)$. However, every monotonic function is differentiable almost everywhere, \emph{i.e.,} the set of points in which $\sigma(x)$ is not differentiable has Lebesgue measure zero.
	We define $\sigma'(x)$ to be equal to zero wherever $\sigma(x)$ is not differentiable; and we take $\sigma'(x)$ to be the derivative of $\sigma(x)$ wherever it is differentiable.
	With this definition of $\sigma'(x)$ and from the continuity of $\sigma(x)$, we have that the integral of $\sigma'(x)/\sigma(x)$ gives us back the function $\log(\sigma(x))$.

	Since $\sigma(x)$ is an increasing positive function, and $Z\geq\delta>0$, we conclude that
	\begin{equation}
		\E{\log\frac{1+Z\sigma'(X)}{\sigma(X)}\Big|Z\geq\delta}
		\geq \E{\log\frac{1+\delta \sigma'(X)}{\sigma(X)}}.
	\end{equation}
	From Lemma \ref{lmm:hX+logf=hgX} and the fact that the integral of $\sigma'(x)/\sigma(x)$ gives us back the function $\log(\sigma(x))$, we obtain that
	\begin{equation*}
		\h{X} + \E{\log\frac{1+\delta \sigma'(X)}{\sigma(X)}}
		=\h{\psi(X)},
	\end{equation*}
	where $\psi(x)$ is defined in Theorem \ref{thm:1/f+logCapLB}. As a result from \eqref{eqn:IXY|Z}, we obtain that
\if@twocolumn
	\begin{align}
		\I{X}{Y|Z\geq\delta}\geq&\h{Y|Z\geq\delta}-\h{X} \nonumber\\
		&-\E{\log(1+Z\sigma'(X))|Z\geq\delta} \nonumber\\
		&+\h{\psi(X)}-\h{Z|Z\geq\delta}, \label{eqn:IXY|Z-am}
	\end{align}
\else
	\begin{align}
		\I{X}{Y|Z\geq\delta}\geq&\h{Y|Z\geq\delta}-\h{X}
		-\E{\log(1+Z\sigma'(X))|Z\geq\delta} \nonumber\\
		&+\h{\psi(X)}-\h{Z|Z\geq\delta}, \label{eqn:IXY|Z-am}
	\end{align}
\fi
	Using this inequality in conjunction with \eqref{eqn:prf:thm:1/f+logCapLB:I>I|Z>d}, we obtain a lower bound on $\I{X}{Y}$.
	The lower bound that we would like to prove in the statement of the theorem is that 
	\begin{equation*}
		\I{X}{Y}
		\geq \alpha \h{\psi(X)} - \alpha\h{Z|Z\geq\delta}-\mathrm{H}_2(\alpha).
	\end{equation*}
	As a result, it suffices to prove that for all continuous random variables $X$ with pdf $f_X(x)$ we have
	\begin{equation*}
		\h{Y|Z\geq\delta}-\h{X}-\E{\log(1+Z\sigma'(X))|Z\geq\delta}
		\geq 0.
	\end{equation*}
	To this end, observe that $\h{Y|Z\geq\delta}\geq\h{Y|Z,Z\geq\delta}$.
	Thus, if we show that
\if@twocolumn
	\begin{align}
		&\h{Y|Z,Z\geq\delta} \nonumber\\
		&\qquad=\h{X}+\E{\log(1+Z\sigma'(X))|Z\geq\delta}, \label{eqn:hY|Z>hX}
	\end{align}
\else
	\begin{equation} \label{eqn:hY|Z>hX}
		\h{Y|Z,Z\geq\delta}
		=\h{X}+\E{\log(1+Z\sigma'(X))|Z\geq\delta},
	\end{equation}
\fi
	the proof is complete.
	We can write that
	\begin{equation*}
		\h{Y|Z,Z\geq\delta}=\int_\delta^{\infty}{f_{Z'}(z)\h{Y|Z=z}\ud z},
	\end{equation*}
	where $Z'$ is $Z$ conditioned to $Z\geq\delta$, and the pdf of $Z'$ is denoted by $f_{Z'}(z)$.
	By defining the function $r_z(x):= x+z \sigma(x)$, we obtain that
	\begin{align*}
		Y_z=r_z(X),
	\end{align*}
	where $Y_z$ is $Y$ which is conditioned to $Z=z\geq\delta$.
	Since, $\sigma(x)$ is a continuous increasing function, $r_z(x)$ is a bijection for all $z\geq\delta$, and so its inverse function, $r_z^{-1}(y)$, exists.
	Moreover, since $X$ is continuous and $r_z(\cdot)$ is a bijection, $Y_z$ is also continuous random variable with pdf $f_Y(y)$ defined as following:
	\begin{equation*}
		f_{Y_z}(y) = \frac{1}{1+z\sigma'(x)} f_X(x),
	\end{equation*}
	where $x=r_z^{-1}(y)$.
	\footnote{The  measure zero points where $\sigma(x)$ is not differentiable affect $f_{Y_z}(y)$ on a measure zero points. However, note that $F_{Y_z}(y)=F_X(r^{-1}(y))$ is always correct and thus the values of $f_{Y_z}(y)$ on a measure zero set of points are not important.}
	Thus, we have that
	\begin{align*}
		\h{Y|Z=z}=&\E{\log\frac{1}{f_{Y_z}(Y_z)}}\\
		=&\E{\log\frac{1}{f_{X}(X)}}+\E{\log(1+z \sigma'(X))}\\
		=&\h{X}+\E{\log(1+z \sigma'(X))}.
	\end{align*}
	By taking expected value over $Z\geq\delta$ from both sides, \eqref{eqn:hY|Z>hX} is achieved.
	Therefore, the theorem is proved.
\qed

%%%%%%%%%%%%%%%%%%%%%%
% Proof of Theorem: KL Symmetric Upper Bound
\subsection{Proof of Theorem \ref{thm:KLUpBnd}} \label{subsec:Proof:thm:KLUpBnd}
	Based on \cite{aminian2015capacity} we obtain that
	\begin{equation*}
		\I{X}{Y}
		\leq\mathrm{D}_{\text{sym}}(\mu_{X,Y} \| \mu_X\mu_Y),
	\end{equation*}
	Utilizing Lemma \ref{lmm:hY|X}, we obtain that the pdfs $f_Y(y)$ and $f_{Y|X}(y|x)$ exist and are well-defined.
	Therefore,
\if@twocolumn
	\begin{align*}
		&\mathrm{D}_{\text{sym}}(\mu_{X,Y} \| \mu_X\mu_Y) \\
		&\quad=\mathrm{D}(\mu_{X,Y} \| \mu_X\mu_Y)+\mathrm{D}(\mu_X\mu_Y\|\mu_{X,Y}) \\
		&\quad=\Ex{\mu_{X,Y}}{\log\frac{f_{Y|X}(Y|X)}{f_Y(Y)}}
		+\Ex{\mu_X \mu_Y}{\log\frac{f_Y(Y)}{f_{Y|X}(Y|X)}} \\
		&\quad=-\Ex{\mu_{X,Y}}{\log\frac{1}{f_{Y|X}(Y|X)}}
		+\E{\log\frac{1}{f_Y(Y)}} \\
		&\qquad+\Ex{\mu_X\mu_Y}{\log\frac{1}{f_{Y|X}(Y|X)}}
		-\E{\log\frac{1}{f_Y(Y)}}\\
		&\quad=\Ex{\mu_X\mu_Y}{\log\frac{1}{f_{Y|X}(Y|X)}}-\h{Y|X}.
	\end{align*}
\else
	\begin{align*}
		\mathrm{D}_{\text{sym}}(\mu_{X,Y} \| \mu_X\mu_Y)
		=&\mathrm{D}(\mu_{X,Y} \| \mu_X\mu_Y)+\mathrm{D}(\mu_X\mu_Y\|\mu_{X,Y}) \\
		=&\Ex{\mu_{X,Y}}{\log\frac{f_{Y|X}(Y|X)}{f_Y(Y)}}
		+\Ex{\mu_X \mu_Y}{\log\frac{f_Y(Y)}{f_{Y|X}(Y|X)}} \\
		=&-\Ex{\mu_{X,Y}}{\log\frac{1}{f_{Y|X}(Y|X)}}
		+\E{\log\frac{1}{f_Y(Y)}} \\
		&+\Ex{\mu_X\mu_Y}{\log\frac{1}{f_{Y|X}(Y|X)}}
		-\E{\log\frac{1}{f_Y(Y)}}\\
		=&\Ex{\mu_X\mu_Y}{\log\frac{1}{f_{Y|X}(Y|X)}}-\h{Y|X}.
	\end{align*}
\fi
	Again, from Lemma \ref{lmm:hY|X}, since $Z\sim\mathcal{N}(0,1)$, we obtain that
	\begin{equation*}
		\log{\frac{1}{f_{Y|X}(y|x)}}
		=\log\left(\sigma(x)\sqrt{2\pi}\right)
		+\frac{(y-x)^2}{2\sigma^2(x)}
	\end{equation*}
	Therefore, since $Z=(Y-X)/\sigma(X)$, we obtain that
	\begin{equation} \label{eqnhY|X12}
		\h{Y|X} = \E{\log\left(\sigma(X)\sqrt{2\pi}\right)} + \frac{1}{2}. 
	\end{equation}
	In addition,
\if@twocolumn
	\begin{align*}
		&\Ex{\mu_X\mu_Y}{\log\frac{1}{f_{Y|X}(Y|X)}} \\
		&\qquad=\E{\log\left(\sqrt{2\pi}\sigma(X)\right)}
		+\Ex{\mu_X\mu_Y}{\frac{(Y-X)^2}{2 \sigma^2(X)}}.
	\end{align*}
\else
	\begin{equation*}
		\Ex{\mu_X\mu_Y}{\log\frac{1}{f_{Y|X}(Y|X)}}
		=\E{\log\left(\sqrt{2\pi}\sigma(X)\right)}
		+\Ex{\mu_X\mu_Y}{\frac{(Y-X)^2}{2 \sigma^2(X)}}.
	\end{equation*}
\fi
	By expanding, we obtain that
\if@twocolumn
	\begin{align*}
		\Ex{\mu_X\mu_Y}{\frac{(Y-X)^2}{\sigma^2(X)}}
		=&\E{Y^2}\E{\frac{1}{\sigma^2(X)}}
		+\E{\frac{X^2}{\sigma^2(X)}} \\
		&-2\E{Y}\E{\frac{X}{\sigma^2(X)}}.
	\end{align*}
\else
	\begin{equation*}
		\Ex{\mu_X\mu_Y}{\frac{(Y-X)^2}{\sigma^2(X)}}
		=\E{Y^2}\E{\frac{1}{\sigma^2(X)}}
		+\E{\frac{X^2}{\sigma^2(X)}}
		-2\E{Y}\E{\frac{X}{\sigma^2(X)}}.
	\end{equation*}
\fi
	By substituting $Y$ with $X+\sigma(X) Z$ and simplifying we can write
\if@twocolumn
	\begin{align*}
		&\Ex{\mu_X\mu_Y}{\frac{(Y-X)^2}{\sigma^2(X)}} \\
		&\qquad=\E{X^2}\E{\frac{1}{\sigma^2(X)}}
		+\E{\sigma^2(X)}\E{\frac{1}{\sigma^2(X)}} \\
		&\qquad\quad+\E{\frac{X^2}{\sigma^2(X)}}
		-2\E{X}\E{\frac{X}{\sigma^2(X)}} \\
		&\qquad=\bigg(\E{X^2}\E{\frac{1}{\sigma^2(X)}}
		+\E{\sigma^2(X)}\E{\frac{1}{\sigma^2(X)}} \\
		&\qquad\quad-\E{\frac{X^2+\sigma^2(X)}{\sigma^2(X)}} \bigg)
		+1 \\
		&\qquad\quad +2\left(\E{\frac{X^2}{\sigma^2(X)}}
		-2\E{X}\E{\frac{X}{\sigma^2(X)}}\right),
	\end{align*}
\else
	\begin{align*}
		\Ex{\mu_X\mu_Y}{\frac{(Y-X)^2}{\sigma^2(X)}}
		=&\E{X^2}\E{\frac{1}{\sigma^2(X)}}
		+\E{\sigma^2(X)}\E{\frac{1}{\sigma^2(X)}} \\
		&+\E{\frac{X^2}{\sigma^2(X)}}
		-2\E{X}\E{\frac{X}{\sigma^2(X)}} \\
		=&\E{X^2}\E{\frac{1}{\sigma^2(X)}}
		+\E{\sigma^2(X)}\E{\frac{1}{\sigma^2(X)}}
		-\E{\frac{X^2+\sigma^2(X)}{\sigma^2(X)}}
		+1\\
		&+2\left(\E{\frac{X^2}{\sigma^2(X)}}
		-2\E{X}\E{\frac{X}{\sigma^2(X)}}\right),
	\end{align*}
\fi
	which equals to
	\begin{align*}
		1-\Cov{X^2+\sigma^2(X)}{\frac{1}{\sigma^2(X)}}
		+2\Cov{X}{\frac{X}{\sigma^2(X)}}.
	\end{align*}
	Therefore, from all above equations the theorem is proved. 
\qed

%%%%%%%%%%%%%%%%%%%%%%%%%%
% Proof of Corollary: KL Capacity Symmetric Upper Bound

\subsection{Proof of Corollary \ref{cor:KLCapUpBnd}} \label{subsec:Proof:cor:KLCapUpBnd}
	Observe that
	\begin{align*}
		\frac 12 F&=\frac12 \left(\frac{u^2}{\sigma^2(u)}
		+\frac{u^2}{\sigma^2(0)}
		+\frac{\sigma^2(0)}{\sigma^2(u)}
		+\frac{\sigma^2(u)}{\sigma^2(0)}
		-2\right)\\
		&=\frac{1}{2}\left(u^2+\sigma^2(u)-\sigma^2(0) \right)
		\left(\frac{1}{\sigma^2(0)}-\frac{1}{\sigma^2(u)} \right)
+\frac{u^2}{\sigma^2(u)}.
	\end{align*}
	Then, using Theorem \ref{thm:KLUpBnd},  it suffices to prove the following two inequalities:
\if@twocolumn
	\begin{align}
		&\Cov{X^2+\sigma^2(X)}{\frac{-1}{\sigma^2(X)}} \label{eqn:KLCapUpBndCov1}\\
		&\qquad\leq \beta
		\left(u^2+\sigma^2(u)-\sigma^2(0) \right)
		\left(\frac{1}{\sigma^2(0)}-\frac{1}{\sigma^2(u)} \right), \nonumber
	\end{align}
\else
	\begin{equation} \label{eqn:KLCapUpBndCov1}
		\Cov{X^2+\sigma^2(X)}{\frac{-1}{\sigma^2(X)}}
		\leq \beta
		\left(u^2+\sigma^2(u)-\sigma^2(0) \right)
		\left(\frac{1}{\sigma^2(0)}-\frac{1}{\sigma^2(u)} \right),
	\end{equation}
\fi
and 
	\begin{equation} \label{eqn:KLCapUpBndCov2}
		\Cov{X}{\frac{X}{\sigma^2(X)}}
		\leq \beta\frac{u^2}{\sigma^2(u)},
	\end{equation}
	where
	\begin{equation*}
		\beta=
		\begin{cases}
			\frac{1}{4} & \alpha \geq \frac{u}{2} \\
			\frac{u}{\alpha}\left(1-\frac{u}{\alpha}\right) & \alpha < \frac{u}{2}
		\end{cases}.
	\end{equation*}
	Since $\sigma(x)$ is increasing, we obtain that $x^2+\sigma^2(x)$ and $-1/\sigma^2(x)$ are also increasing.
	Therefore, from Lemma \ref{lmm:ConvCover}, equation \eqref{eqn:KLCapUpBndCov1} is proved.
	Similarly, \eqref{eqn:KLCapUpBndCov2} is also obtained from Lemma \ref{lmm:ConvCover} because $x$ and $x/\sigma^2(x)$ are increasing functions.
\qed

%%%%%%%%%%%%%%%%%%%%
% Proof of Lemma: Majorization h(Y)>h(X)
\subsection{Proof of Lemma \ref{thm:Maj hY>hX}} \label{subsec:Proof:thm:Maj hY>hX}
	From Definition \ref{def:Ent}, we obtain that
	\begin{equation*}
		\h{X}-\h{Y}
		=\E{\log\frac{f_Y(Y)}{f_X(X)}}.
	\end{equation*}
	Now, utilizing the inequality $\log x \leq x-1$, it suffuces to prove that
	\begin{equation*}
		\E{\frac{f_Y(Y)}{f_X(X)}} \leq 1.
	\end{equation*}
	To this end, we can write
	\begin{align*}
		\E{\frac{f_Y(Y)}{f_X(X)}}
		=&\int_\mathcal{X}\int_\mathcal{Y}{f_{X,Y}(x,y) \frac{f_Y(y)}{f_X(x)} \ud y \ud x} \\
		=&\int_\mathcal{X}\int_\mathcal{Y}{f_{Y|X}(y|x) f_Y(y) \ud y \ud x} \\
		=&\int_\mathcal{Y}{f_Y(y) \int_\mathcal{X}{f_{Y|X}(y|x)\ud x} \ud y} \\
		\leq&\int_\mathcal{Y}{f_Y(y) \ud y} = 1,
	\end{align*}
	where the last inequality holds because of the assumption of the lemma.
	Therefore, the lemma is proved.
\qed

%%%%%%%%%%%%
% Proof of Lemma: h(Y|X)
\subsection{Proof of Lemma \ref{lmm:hY|X}} \label{subsec:prf:lmm:hY|X}
	The conditional pdf $f_{Y|X}(y|x)$ can be easily obtained from the definition of channel in \eqref{eqn:ChannelModel}.
	In order to calculate $\h{Y|X}$, using the Definition \ref{def:Ent} we can write
\if@twocolumn
	\begin{align*}
		\h{Y|X}
		=& \E{\frac{1}{f_{Y|X}(Y|X)}} \\
		=& \E{\log \sigma(X)} + \E{\log \frac{1}{f_Z\left(\frac{Y-X}{\sigma(X)}\right)}}.
	\end{align*}
\else
	\begin{equation*}
		\h{Y|X}
		=\E{\frac{1}{f_{Y|X}(Y|X)}}
		=\E{\log \sigma(X)} + \E{\log \frac{1}{f_Z\left(\frac{Y-X}{\sigma(X)}\right)}}.
	\end{equation*}
\fi
	Exploiting the fact that $(Y-X)/\sigma(X)=Z$, $\h{Y|X}$ is obtained.

	It only remains to prove that $Y$ is continuous.
	To this end, from the definition of the channel in \eqref{eqn:ChannelModel}, we obtain that
	\begin{align*}
		F_Y(y) = \Pr{Z\leq\frac{y-X}{\sigma(X)}}
		=\mathbb{E}_{\mu_X}\left[{F_Z \left( \frac{y-X}{\sigma(X)} \right)}\right],
	\end{align*}
	where $F_Y(y)$ and $F_Z(z)$ are the cdfs of the random variables $Y$ and $Z$, defined by $F_Y(y)=\Pr{Y\leq y}$ and $F_Z(z)=\Pr{Z\leq z}$, respectively.
	In order to prove the claim about $f_Y(y)$, we must show that
	\begin{equation*}
		\int_{-\infty}^y{\E{\frac{1}{\sigma(x)}f_Z \left(\frac{y-x}{\sigma(x)} \right)}\ud y}
		=\E{F_Z \left( \frac{y-X}{\sigma(X)} \right)},
	\end{equation*}
	for all $y\in\bR$.
	Because of the Fubini's theorem \cite[Chapter 2.3]{Stein05}, it is equivalent to
	\begin{equation*}
		\lim_{n\to\infty}\E{F_Z\left(\frac{-n-X}{\sigma(X)} \right)} = 0.
	\end{equation*}
	Equivalently, we need to show that for any $\epsilon>0$, there exists $m$ such that
	\begin{equation} \label{eqn:FZn<e}
		\E{F_Z\left(\frac{-n-X}{\sigma(X)} \right)} \leq \epsilon,
		\qquad\forall n>m.
	\end{equation}
	Since $\lim_{z\to-\infty}F_Z(z) = 0$, there exists $\ell\in\bR$ such that
	\begin{equation*}
		F_Z(z) \leq \frac{\epsilon}{2},
		\qquad\forall z\leq \ell.
	\end{equation*}
	Therefore, since $F_Z(z)\leq 1$ for all $z$, we can write
	\begin{equation*}
		\E{F_Z\left(\frac{-n-X}{\sigma(X)} \right)}
		\leq \frac{\epsilon}{2}+\Pr{\frac{-n-X}{\sigma(X)}\geq \ell}.
	\end{equation*}
	We can write
\if@twocolumn
	\begin{align*}
		\Pr{\frac{-n-X}{\sigma(X)}\geq\ell}
		=&\Pr{n\leq-X-\ell\sigma(X)} \\
		\leq& \Pr{n\leq-X}.
	\end{align*}
\else
	\begin{equation*}
		\Pr{\frac{-n-X}{\sigma(X)}\geq\ell}
		=\Pr{X+\ell\sigma(X) \leq -n}.
	\end{equation*}
\fi
	Now, we can take $m$ large enough such that, 
	\begin{equation*}
		\Pr{\frac{-n-X}{\sigma(X)}\geq\ell} \leq \frac{\epsilon}{2},
		\qquad n>m.
	\end{equation*}
	As a result, \eqref{eqn:FZn<e} is proved.
\qed

%%%%%%%%%%%%%%%%%%%%%
% Proof of Lemma: h(X)+E[ln f_{\sigma}(x)]
\subsection{Proof of Lemma \ref{lmm:hX+logf=hgX}} \label{subsec:prf:lmm:hX+logf=hgX}
	Since $\sigma(x)$ is Riemann integrable, $\varphi(x)$ is continuous and since $\sigma(x)>0$ (a.e.), $\varphi(x)$ is a strictly increasing function over the support of $X$.
	It yields that $\varphi(x)$ is an injective function and there exists an inverse function $\varphi^{-1}(\cdot)$ for $\varphi(\cdot)$.
	Now, define random variable $Y=\varphi(X)$.
	Assume that the pdf of $X$ is $f_X(x)$.
	Since $X$ is a continuous random variable and $\varphi(x)$ is a bijection, $Y$ is also continuous random variable with the following pdf:
	\begin{equation*}
		f_Y(y)=\frac{1}{\frac{\ud}{\ud x}\varphi(x)} f_X(x),
	\end{equation*}
	where $x=\varphi^{-1}(y)$.
	Hence, we have that
	\begin{equation*}
		f_Y(y) = \frac{1}{\sigma\left(\varphi^{-1}(y)\right)} f_X\left(\varphi^{-1}(y)\right).
	\end{equation*}
	Now, we can calculate the differential entropy of $Y$ as following:
	\begin{align*}
		\h{Y}=&\E{\log\frac{1}{f_Y(Y)}}\\
		=&\E{\log\frac{\sigma\left(\varphi^{-1}(Y)\right)}{f_X\left(\varphi^{-1}(Y)\right)}}\\
		=&\E{\log\frac{\sigma(X)}{f_X(X)}}\\
		=&\h{X}+\E{\log{\varphi(X)}}.
	\end{align*}
	Therefore, the lemma is proved.
\qed

%%%%%%%%%%%%%%%%%%%%%%
% Proof of Lemma: Using Convex Cover Method
\subsection{Proof of Lemma \ref{lmm:ConvCover}} \label{subsec:Proof:lmm:ConvCover}
	First, assume that $v(x)=ax+b$, with $a>0$.
	We will prove the general case later.
	In this case, we claim that the support of the optimal solution only needs to have two members.
	To this end, note that the following problem is equivalent to the original problem defined in \eqref{eqn:OrigProbCov}:
	\begin{equation*}
		\max_{\gamma\leq\alpha}\max_
		{\substack{\mu_X:\ell\leq X\leq u\\ \E{X}=\gamma}}
		\Cov{w(X)}{v(X)}.
	\end{equation*}
	Since $v(x)=ax+b$, for a given $\gamma$, we would like to maximize
	\begin{align*}
		\Cov{w(X)}{v(X)} = \E{w(X) v(X)}-(a\gamma+b)\E{w(X)},
	\end{align*}
	which is a linear function of $\mu_X$, subject to $\E{X}=\gamma$ which is also a linear function of $\mu_X$.
	By the standard cardinality reduction technique (Fenchel's extension of the Caratheodory theorem), we can reduce the support of $\mu_X$ to at most two members (see \cite[Appendix C]{ElGamalKim} for a discussion of the technique).
	Assume that the support of $\mu_X$ is $\{x_1, x_2\}$ where $\ell \leq x_1 \leq x_2 \leq u$ with pmf $p_X(x_1)=1-p_X(x_2)=p$.
	Thus, we can simplify $\Cov{w(X)}{v(X)}$ as
\if@twocolumn
	\begin{align*}
		&\Cov{w(X)}{v(X)} \\
		&\qquad=\sum_{i=1}^2{p_X(x_i) w(x_i) v(x_i)} \\
		&\qquad\quad-\sum_{i=1}^2\sum_{j=1}^2{p_X(x_i) p_X(x_j) w(x_i) v(x_j)} \\
		&\qquad=p (1-p) \left(w(x_2) - w(x_1)\right) \left(v(x_2) - v(x_1)\right),
	\end{align*}
\else
	\begin{align*}
		\Cov{w(X)}{v(X)}
		=&\sum_{i=1}^2{p_X(x_i) w(x_i) v(x_i)}
		-\sum_{i=1}^2\sum_{j=1}^2{p_X(x_i) p_X(x_j) w(x_i) v(x_j)} \\
		=&p (1-p) \left(w(x_2) - w(x_1)\right) \left(v(x_2) - v(x_1)\right),
	\end{align*}
\fi
	where the last equality can be obtained by expanding the sums.
	Thus, the problem defined in \eqref{eqn:OrigProbCov} equals the following:
	\begin{equation*}
		\max_{\substack
		{p,x_1,x_2:0\leq p\leq 1\\ \ell\leq x_1\leq x_2 \leq u\\p x_1+(1-p) x_2\leq\alpha}}
		p(1-p) \left(w(x_2)-w(x_1)\right) \left(v(x_2)-v(x_1)\right).
	\end{equation*}
	We claim that the optimal choice for $x_1$ is $x_1=\ell$.
	To see this, observe that  $w(x)$ and $v(x)$ are increasing functions, and hence 
	\begin{align*}
		&p \ell +(1-p) x_2 \leq p x_1+(1-p) x_2 \leq \alpha
	\end{align*}
	and
\if@twocolumn
	\begin{align*}
		&\left(w(x_2)-w(\ell)\right)\left(v(x_2)-v(\ell)\right) \\
		&\qquad\geq\left(w(x_2)-w(x_1)\right) \left(v(x_2)-v(x_1)\right).
	\end{align*}
\else
	\begin{equation*}
		\left(w(x_2)-w(\ell)\right)\left(v(x_2)-v(\ell)\right)
		\geq\left(w(x_2)-w(x_1)\right) \left(v(x_2)-v(x_1)\right).
	\end{equation*}
\fi
	Hence, $x_1=\ell$ is optimal.
	Substituting $v(x)=ax+b$, we obtain that the problem is equivalent with the following:
	\begin{equation*}
		a\max_{\substack
		{p,x:0\leq p\leq 1\\ \ell\leq x \leq u\\p \ell+(1-p) x\leq\alpha}}
		p(1-p) \left(w(x)-w(\ell)\right) \left(x-\ell\right).
	\end{equation*}
	Utilizing KKT conditions, one obtains that the optimal solution is
	\begin{align*}
		\begin{cases}
			p^*=\frac{1}{2}, x_1^*=\ell, x_2^* = u & \alpha\geq\frac{\ell+u}{2} \\
			p^*=\frac{u-\alpha}{u-\ell}, x_1^*=\ell, x_2^* =u & \alpha<\frac{\ell+u}{2} 
		\end{cases}.
	\end{align*}

	Now, we consider the general case of $v(x)$ being a convex function (but not necessarily linear).
	Since $v(x)$ is convex, we obtain that
	\begin{equation*}
		v(x) \leq v(\ell) + (x-\ell) \frac{v(u)-v(\ell)}{u-\ell},
		\qquad \forall x\in [\ell,u].
	\end{equation*}
	The right hand side is the line that connects the two points $(\ell, v(\ell))$ and $(u, v(u))$; this line lies above the curve $x\mapsto v(x)$ for any $x\in [\ell, u]$. 
	Therefore,
	\begin{equation*}
		\mathbb{E}[v(X)] \leq v(\ell) + (\mathbb{E}[X]-\ell) \frac{v(u)-v(\ell)}{u-\ell}.
	\end{equation*}
	Thus, $\E{X}\leq\alpha$ implies that $\E{v(X)}\leq \Delta$ where
	\begin{equation*}
		\Delta=v(\ell) + (\alpha-\ell) \frac{v(u)-v(\ell)}{u-\ell}.
	\end{equation*}
	Now, we relax the optimization problem and consider 
	\begin{equation*}
		\max_{\substack{\mu_X:\ell\leq X\leq u\\ \E{v(X)}\leq\Delta}}
		\Cov{w(X)}{v(X)}.
	\end{equation*}
	The solution of the above optimization problem is an upper bound for the original problem because the feasible set of the original problem is a subset of the feasible set of the relaxed optimization problem.

	Now, using similar ideas as in the linear case, we conclude that the support of the optimal $\mu_X$ has at most two members.
	And the optimal solution is
	\begin{equation*}
		\begin{cases}
			p^*=\frac{1}{2}, x_1^*=\ell, x_2^* = u & \alpha\geq\frac{\ell+u}{2} \\
			p^*=\frac{v(u)-\Delta}{v(u)-v(\ell)}, x_1^*=\ell, x_2^*=u & \alpha<\frac{\ell+u}{2} 
		\end{cases}.
	\end{equation*}
	It can be verified that
	\begin{equation*}
		\frac{v(u)-\Delta}{v(u)-v(\ell)}=\frac{u-\alpha}{u-\ell}.
	\end{equation*}
	Note that in the case $\alpha>(\ell+u)/2$, we obtain that $\E{X^*}=(\ell+u)/2<\alpha$, where $X^*$ distributed with the optimal probability measure.
	As a result the constraint $\E{X}\leq\alpha$ is redundant.
	Therefore, the support of the optimal $\mu_X$ has two members, which shows that the upper bound is tight in this case.
\qed

%%%%%%%
% Conclusion
%%%%%%%
\section{Conclusion}
In this paper, we studied the capacity of a class of signal-dependent additive noise channels.
These channels are of importance in molecular and optical communication; we also gave a number of new application of such channels in the introduction.
A set of necessary and a set of sufficient conditions for finiteness of capacity were given.
We then introduced two new techniques for proving explicit lower bounds on the capacity.
As a result, we obtained two lower bounds on the capacity.
These lower bounds were helpful in inspecting when channel capacity becomes infinity.
We also provided an upper bound using the symmetrized KL divergence bound.

%%%%%%%
% Bibliography
%%%%%%%
\bibliographystyle{ieeetr}
\bibliography{../refnew}

%%%%%%
% Appendix
%%%%%%
\appendix
\section{Proof of Equation \eqref{eqn:integral-erf}} \label{Appendix:integral-erf}
	Take some arbitrary $x>0$. Then, equation \eqref{eqn:integral-erf} holds because
\if@twocolumn
	\begin{align*}
		&\int_{0}^{\infty}\frac{\sqrt{2}}{\sqrt{\pi c_1^2 }}\ue^{\frac{-(y-v^2)^2}{2c_1^2v^2}} \ud v \\
		&\quad=\int_{0}^{\infty}\frac{1}{\sqrt{\pi}}\left(\frac{v^2+y}{v^2\sqrt{2c_1^2}}
		+\frac{v^2-y}{v^2\sqrt{2c_1^2}}\right)
		\ue^{\frac{-(y-v^2)^2}{2c_1^2v^2}} \ud v \\
		&\quad=\ue^{\frac{2y}{c_1^2}}
		\int_{0}^{\infty}\frac{1}{\sqrt{\pi}}\frac{v^2-y}{v^2\sqrt{2c_1^2}}
		\ue^{\frac{-(y+v^2)^2}{2c_1^2v^2}} \ud v \\
		&\qquad-\int_{0}^{\infty}\frac{1}{\sqrt{\pi}}\frac{-(v^2+y)}{v^2\sqrt{2c_1^2}}
		\ue^{\frac{-(y-v^2)^2}{2c_1^2v^2}} \ud v \\
		&\quad=\ue^{\frac{2y}{c_1^2}}
		\int_{0}^{x}\frac{1}{\sqrt{\pi}}\frac{v^2-y}{v^2\sqrt{2c_1^2}}
		\ue^{\frac{-(y+v^2)^2}{2c_1^2v^2}} \ud v \\
		&\qquad+\ue^{\frac{2y}{c_1^2}}
		\int_{x}^{\infty}\frac{1}{\sqrt{\pi }}\frac{v^2-y}{v^2\sqrt{2c_1^2}}
		\ue^{\frac{-(y+v^2)^2}{2c_1^2v^2}} \ud v \\
		&\qquad-\int_{0}^{x}\frac{1}{\sqrt{\pi}}\frac{-(v^2+y)}{v^2\sqrt{2c_1^2}}
		\ue^{\frac{-(y-v^2)^2}{2c_1^2v^2}} \ud v \\
		&\qquad-\int_{x}^{\infty}\frac{1}{\sqrt{\pi}}\frac{-(v^2+y)}{v^2\sqrt{2c_1^2}}
		\ue^{\frac{-(y-v^2)^2}{2c_1^2v^2}} \ud v.
	\end{align*}
\else
	\begin{align*}
		\int_{0}^{\infty}
		{\frac{\sqrt{2}}{\sqrt{\pi c^2 }}
		\ue^{-\frac{(y-v^2)^2}{2 c^2 v^2}} \ud v}
		=&\int_{0}^{\infty}
		{\frac{1}{\sqrt{\pi}}
		\left(\frac{v^2+y}{v^2 \sqrt{2 c^2}}+\frac{v^2-y}{v^2\sqrt{2 c^2}}\right)
		\ue^{\frac{-(y-v^2)^2}{2 c^2v^2}} \ud v} \\
		=&\ue^{\frac{2y}{c^2}}
		\int_{0}^{\infty}
		{\frac{1}{\sqrt{\pi}}\frac{v^2-y}{v^2\sqrt{2 c^2}}
		\ue^{\frac{-(y+v^2)^2}{2 c^2v^2}} \ud v}
		+\int_{0}^{\infty}
		{\frac{1}{\sqrt{\pi}}\frac{v^2+y}{v^2\sqrt{2 c^2}}
		\ue^{\frac{-(y-v^2)^2}{2 c^2 v^2}} \ud v} \\
		=&\ue^{\frac{2y}{c^2}}
		\int_{0}^{1}
		{\frac{1}{\sqrt{\pi}}\frac{v^2-y}{v^2\sqrt{2 c^2}}
		\ue^{\frac{-(y+v^2)^2}{2 c^2 v^2}} \ud v}
		+\ue^{\frac{2y}{c^2}}
		\int_{1}^{\infty}
		{\frac{1}{\sqrt{\pi}}\frac{v^2-y}{v^2\sqrt{2 c^2}}
		\ue^{\frac{-(y+v^2)^2}{2 c^2 v^2}} \ud v} \\
		&+\int_{0}^{1}
		{\frac{1}{\sqrt{\pi}}\frac{v^2+y}{v^2\sqrt{2 c^2}}
		\ue^{\frac{-(y-v^2)^2}{2 c^2v^2}} \ud v}
		+\int_{1}^{\infty}
		{\frac{1}{\sqrt{\pi}}\frac{v^2+y}{v^2\sqrt{2 c^2}}
		\ue^{\frac{-(y-v^2)^2}{2 c^2 v^2}} \ud v}.
	\end{align*}
\fi
	Now utilize the change of variables
	\begin{equation*}
		v \mapsto u_1=\frac{y-v^2}{v\sqrt{2 c^2}},
		\qquad v \mapsto u_2=\frac{y+v^2}{v\sqrt{2 c^2}}
	\end{equation*}
	to re-express the above integrals.
	Note that
	\begin{equation*}
		\ud u_1 = -\frac{v^2+y}{v^2\sqrt{2 c^2}} \ud v,
		\qquad \ud u_2= \frac{v^2-y}{v^2\sqrt{2 c^2}} \ud v.
	\end{equation*}
	For $y>0$, if $v=0$ then $u_1=+\infty$, $u_2=+\infty$, if $v=+\infty$ then $u_1=-\infty$, $u_2=+\infty$ and if $v=1$ then $u_1=(y-1)/\sqrt{2 c^2}$, $u_2=(y+1)/\sqrt{2 c^2}$.
	For $y<0$, if $v=0$ then $u_1=-\infty$, $u_2=-\infty$, if $v=+\infty$ then $u_1=-\infty$, $u_2=+\infty$ and if $v=1$ then $u_1=(y-1)/\sqrt{2 c^2}$, $u_2=(y+1)/\sqrt{2 c^2}$.
	Now for $y>0$ we have
\if@twocolumn
	\begin{align*}
		&-\frac{\ue^{\frac{2y}{c_1^2}}}{2}
		\int_{\frac{y+x^2}{\sqrt{2c_1^2}x}}^{\infty}
		\frac{2}{\sqrt{\pi}} \ue^{-t_2^2} \ud t_2
		+\frac{\ue^{\frac{2y}{c_1^2}}}{2}
		\int_{\frac{y+x^2}{x\sqrt{2c_1^2}}}^{\infty}
		\frac{2}{\sqrt{\pi }} \ue^{-t_2^2} \ud t_2 \\
		&\qquad+\frac{1}{2}
		\int_{\frac{y-x^2}{x\sqrt{2c_1^2}}}^{\infty}
		\frac{2}{\sqrt{\pi}} \ue^{-t_1^2} \ud t_1
		+\frac{1}{2}
		\int_{-\infty}^{\frac{y-x^2}{x\sqrt{2c_1^2}}}
		\frac{2}{\sqrt{\pi}} \ue^{-t_1^2} \ud t_1 \\
		&\quad=\frac{1}{2}
		\int_{-\infty}^{\infty} \frac{2}{\sqrt{\pi}} \ue^{-t_1^2} \ud t_1 \\
		&\quad=\frac{1}{2} \int_{0}^{\infty}\frac{2}{\sqrt{\pi}} \ue^{-t_1^2} \ud t_1
		+\frac{1}{2} \int_{-\infty}^{0}\frac{2}{\sqrt{\pi}} \ue^{-t_1^2} \ud t_1\\
		&\quad=\frac{1}{2}\textit{erf}(t_1)|_{-\infty}^{\infty}
		=1,
	\end{align*}
\else
	\begin{align*}
		&-\ue^{\frac{2y}{c^2}}
		\int_{\frac{y+1}{\sqrt{2 c^2}}}^{\infty}
		{\frac{1}{\sqrt{\pi}} \ue^{-u_2^2} \ud u_2}
		+\ue^{\frac{2 y}{c^2}}
		\int_{\frac{y+1}{\sqrt{2 c^2}}}^{\infty}
		{\frac{1}{\sqrt{\pi}} \ue^{-u_2^2} \ud u_2}
		+\int_{\frac{y-1}{\sqrt{2 c^2}}}^{\infty}
		{\frac{1}{\sqrt{\pi}} \ue^{-u_1^2} \ud u_1}
		+\int_{-\infty}^{\frac{y-1}{\sqrt{2 c^2}}}
		{\frac{1}{\sqrt{\pi}} \ue^{-u_1^2} \ud u_1} \\
		&\qquad=
		\int_{-\infty}^{\infty}
		{\frac{1}{\sqrt{\pi}} \ue^{-u_1^2} \ud u_1}
		=1.
	\end{align*}
\fi
	Similarly, for $y<0$ we have
\if@twocolumn
	\begin{align*}
		&-\frac{\ue^{\frac{2y}{c_1^2}}}{2}
		\int_{-\infty}^{\frac{y+x^2}{x\sqrt{2c_1^2}}}
		\frac{2}{\sqrt{\pi }} \ue^{-t_2^2} \ud t_2
		+\frac{\ue^{\frac{2y}{c_1^2}}}{2}
		\int_{\frac{y+x^2}{x\sqrt{2c_1^2}}}^{\infty}
		\frac{2}{\sqrt{\pi}} \ue^{-t_2^2} \ud t_2\\
		&\qquad -\frac{1}{2} \int_{-\infty}^{\frac{y-x^2}{x\sqrt{2c_1^2}}}
		\frac{2}{\sqrt{\pi}} \ue^{-t_1^2} \ud t_1
		+\frac{1}{2} \int_{-\infty}^{\frac{y-x^2}{x\sqrt{2c_1^2}}}
		\frac{2}{\sqrt{\pi}} \ue^{-t_1^2} \ud t_1 \\
		&\quad=\frac{\ue^{\frac{2y}{c_1^2}}}{2}\textit{erf}(t_2)|_{-\infty}^{\infty}
		=\ue^{\frac{2y}{c_1^2}},
	\end{align*}
\else
	\begin{align*}
		&\ue^{\frac{2y}{c^2}}
		\int_{-\infty}^{\frac{y+1}{\sqrt{2 c^2}}}
		{\frac{1}{\sqrt{\pi }} \ue^{-u_2^2} \ud u_2}
		+\ue^{\frac{2y}{c^2}}
		\int_{\frac{y+1}{\sqrt{2 c^2}}}^{\infty}
		{\frac{1}{\sqrt{\pi}} \ue^{-u_2^2} \ud u_2}
		-\int_{-\infty}^{\frac{y-1}{\sqrt{2 c^2}}}
		{\frac{1}{\sqrt{\pi}} \ue^{-u_1^2} \ud u_1}
		+\int_{-\infty}^{\frac{y-1}{\sqrt{2 c^2}}}
		{\frac{1}{\sqrt{\pi}} \ue^{-u_1^2} \ud u_1} \\
		&\qquad=\ue^{\frac{2 y}{c^2}}
		\int_{-\infty}^{\infty}
		{\frac{1}{\sqrt{\pi}} \ue^{-u_1^2} \ud u_1}
		=\ue^{\frac{2y}{c^2}}.
	\end{align*}
\fi
	Therefore, the proof is complete.
\qed
\end{document}